\documentclass[11pt]{article}
\pdfoutput=1
\usepackage{amsmath}
\usepackage{amssymb}
\usepackage{cite}
\usepackage{graphicx,bbm,mathrsfs}
\usepackage{nicefrac}
\usepackage{bbm}
\usepackage{geometry}       
\usepackage{jheppub}       

\def\ie{{\it i.e.}}

\newcommand{\be}{\begin{equation}}  
\newcommand{\ee}{\end{equation}}  
\newcommand{\bea}{\begin{eqnarray}}  
\newcommand{\eea}{\end{eqnarray}}

\newcommand{\tr}{\operatorname{tr}}

\renewcommand{\O}{\mathcal O}

\newcommand{\WB}{W\hspace{-1.5pt}B}
\newcommand{\WW}{W\hspace{-.5pt}W}
\newcommand{\BB}{B\hspace{-.5pt}B}
\newcommand{\nn}{\nonumber}

\newcommand\lsim{\mathrel{\rlap{\lower4pt\hbox{\hskip1pt$\sim$}}
    \raise1pt\hbox{$<$}}}
\newcommand\gsim{\mathrel{\rlap{\lower4pt\hbox{\hskip1pt$\sim$}}
    \raise1pt\hbox{$>$}}}

\newcommand{\ferm}{\nicefrac{\mathbf 1}{\mathbf 2}}

\newcommand{\captionfonts}{\small}

\newcommand{\approptoinn}[2]{\mathrel{\vcenter{
  \offinterlineskip\halign{\hfil$##$\cr
    #1\propto\cr\noalign{\kern2pt}#1\sim\cr\noalign{\kern-2pt}}}}}

\makeatletter  
\long\def\@makecaption#1#2{%
  \vskip\abovecaptionskip
  \sbox\@tempboxa{{\captionfonts #1: #2}}%
  \ifdim \wd\@tempboxa >\hsize
    {\captionfonts #1: #2\par}
  \else
    \hbox to\hsize{\hfil\box\@tempboxa\hfil}%
  \fi
  \vskip\belowcaptionskip}
\makeatother   

\title{ Anomalous gauge couplings from composite Higgs and warped extra dimensions}

\author[a]{Sylvain Fichet, }

\affiliation[a]{International Institute of Physics, UFRN, 
Av. Odilon Gomes de Lima, 1722 - Capim~Macio - 59078-400 - Natal-RN, Brazil}

\author[b]{Gero von Gersdorff,}	

\affiliation[b]{ICTP South American Institute for Fundamental Research,
Instituto de Fisica Teorica, Sao Paulo State University, Brazil}

\abstract{
We examine trilinear and quartic anomalous gauge couplings (AGCs) generated in composite Higgs models and models with warped extra dimensions. 
We first  revisit the $SU(2)_L\times U(1)_Y$ effective Lagrangian
and derive the charged and two-photon neutral AGCs.
We derive the general perturbative contributions to the pure field-strength operators from spin $0$, $\frac{1}{2}$, $1$ resonances by means of the heat kernel method. 
In the composite Higgs framework, we derive the pattern of expected deviations from typical $SO(N)$ embeddings of the light composite top partner. 
We then study a generic warped extra dimension framework with $AdS_5$ background, recasting in few parameters the features of models relevant for AGCs.  
We also present a detailed study of the latest bounds from electroweak and Higgs precision observables, with and without brane kinetic terms. For vanishing brane kinetic terms, we find that the $S$ and $T$ parameters exclude KK gauge modes of the RS custodial [non-custodial] scenario  below  $7.7$ [$14.7$] TeV,  for a brane Higgs  and below $6.6$ [$8.1$] TeV for a Pseudo Nambu-Goldstone Higgs, at $95\%$ CL.
These constraints can be relaxed in presence of brane kinetic terms. 
The leading AGCs are probing the KK gravitons and the KK modes of bulk gauge fields in parts of the parameter space.
In these scenarios, the future CMS and ATLAS forward proton detectors could   be sensitive to the effect of   KK gravitons in the multi-TeV mass range. }

\begin{document}

\maketitle

\section{Introduction}
\label{se:intro}

Several major facts like  the gauge-hierarchy problem or the observation of dark matter  suggest that  physics beyond Standard Model of particles (SM) should be present in nature at a mass scale relatively close to the electroweak scale. However, in spite of naturalness arguments, the paradigm of a TeV-scale new physics is challenged by both direct LHC searches and by indirect observables like the LEP electroweak precision tests. 
In a scenario of new physics out of reach from direct observation at the LHC, one may expect that the first manifestations show up in precision measurements of the  SM properties. 

Such precision  tests are already well advanced in the electroweak and flavour sectors of the SM. After the discovery of a new Higgs-like scalar at the LHC,
deviations of  the Higgs sector from its SM predictions are also starting to be scrutinized.
However, during the LHC era,  yet another sector of the SM can be tested to high precision, the one of pure gauge interactions. This is a domain relevant for the CMS and ATLAS main detectors, and also for the future CMS and ATLAS forward proton (FP)  detectors \cite{Tasevsky:2009je,Royon:2013en,Albrow:2013yka} whose construction may start in 2014.

Just like the couplings of other sectors of the SM, anomalous gauge couplings can be systematically analyzed through the effective Lagrangian framework. This approach relies on the  assumption  that the energy scale $E$ accessible by the experiment is  small with respect to the new physics mass scale $\Lambda$ , \be E\ll\Lambda \,. \ee
This fundamental assumption allows one to describe any manifestation of new physics
by local operators of higher dimension. 
 The low-energy effective Lagrangian is written as the SM Lagrangian plus an infinite sum of these higher dimensional operators (HDOs),
\be
\mathcal{L}_{\rm eff}=\mathcal{L}_{\rm SM}+\sum_i \frac{\alpha_i}{\Lambda^{m_i}}\mathcal{O}_i\,.\label{Leff}
\ee 
For a given set  of observables of interest (in our case, the anomalous gauge couplings), it is then sufficient to keep only the  operators at the appropriate first orders of the expansion. By construction these operators capture the leading effects induced by the heavy new physics.

In the present study, the observables of interest are the trilinear gauge couplings and quartic gauge couplings (TGCs, QGCs). As the FP detectors  provide particularly good sensitivity to the process of diffractive photon fusion, we are going to limit our study to two photon QGCs. 
In order to  describe  the effects of new physics on TGCs and QGCs, it turns out that we  have to keep  dimension $6$ operators for charged anomalous gauge couplings (CGCs) and  dimension $8$ operators for neutral anomalous gauge couplings (NGCs). Higher order effects in the observables will be of order $O(E^2/\Lambda^2,v^2/\Lambda^2)$ where $E$ is the typical energy involved in the process. We will also keep some charged dimension $8$ operators relevant for two-photon physics.
Anomalous gauge couplings have already been studied from the effective Lagrangian point of view, see e.g. \cite{Belanger:1999aw, Gupta:2011be}, 
and the recent review \cite{Degrande:2013rea} and references therein.

In this paper we aim at going one step further,  by mapping  actual models of new physics onto the effective Lagrangian. In particular we investigate the effects of composite Higgs  and of warped extra dimensions. Although we consider actual theories of new physics, we adopt an effective parametrization of the theories in order that our predictions cover the broadest range of possibilities and facilitate comparison with other approaches. 

%

The content of this paper is as follows. In Sec.~\ref{se:L_effective}
 we lay out the dimension $8$ effective Lagrangian relevant for TGCs and QGCs in a  basis appropriate to distinguish  between loop-generated and potentially tree-level generated operators. 
  In Sec.~\ref{L_AGC} we derive the anomalous gauge couplings, which are directly probed by experiments. 
  We distinguish two energy regimes,  $E<v$ and $v<E<\Lambda$, in which different set of HDOs become dominant in the experiments.
    In Sec.~\ref{Generic_L_partial} we derive the generic perturbative contributions  from any heavy state to the set of HDOs dominating in the  $v<E<\Lambda$ regime by means of the heat kernel method. The final formulas are straightforwardly applicable to any new physics model. 
    In Sec.~\ref{CHmodels} we discuss and evaluate the effects of composite resonances in   composite Higgs models.
    In Sec.~\ref{ADS5} we study warped extra dimensions,  
    including the case of brane gauge field (sometimes referred to as RS I) and the Randall-Sundrum model with bulk gauge fields, as well as the composite Higgs models via holography.  
     Section \ref{constraints} contains an update on the bounds from Higgs and electrowek precision tests on KK gauge modes and radion masses.     
      In  Sec.~\ref{results} we present and discuss the anomalous couplings predicted in the warped $AdS_5$ framework, including the effect of the radion/dilaton and of the KK gravitons. 
Section \ref{conclu} is devoted to our conclusions.

\section{The effective Lagrangian}
\label{se:L_effective}

First we  define the set of higher dimensional operators entering the effective Lagrangian Eq.~\eqref{Leff}.
\footnote{It is sufficient to use the linearly realized $SU(2)\times U(1)$ Lagrangian, as the nonlinear realization tends to the linear one in the  limit $v\ll\Lambda$. See e.g.  \cite{Brivio:2013pma} for more details about anomalous gauge couplings in the nonlinear realization.  }
 The complete Lagrangian describing anomalous gauge couplings in the broken phase, needed for  phenomenology and experimental studies, will  be derived in the next section. Throughout this paper we use the conventions of \cite{Peskin:1995ev}, including the mostly minus metric signature $(+,-,-,-)$.
We limit the expansion of  $\mathcal{L}_{\rm eff}$ to order $8$, such that
\be
\mathcal{L}_{\rm eff}=\mathcal{L}_{\rm SM}+\mathcal{L}_{(6)}+\mathcal{L}_{(8)}\,.
\ee
The $\mathcal{L}_{(6)}$ piece induces the leading deviations to TGCs, while the $\mathcal{L}_{(8)}$  generates anomalous quartic NGCs. 

The HDOs at a given order can be brought into an irreducible set, \ie~a basis,  using the equivalence relations provided by  integration by parts and equations of motion. There is a freedom in the choice of the particular basis.~\footnote{Note that although the effects on the physical observables from different basis must be the same, they can appear in a rather different way. For example, a modification of the $W$ propagator in a given basis may be recasted in a shift of the Fermi constant in another basis. }
Here our dimension six basis is defined in such a way that (in renormalizable theories) the operators $\mathcal{O}_{FF}$, $\mathcal{O}_{W^3}$ are only generated pertubatively at the loop level while the other ones can potentially be generated at tree-level. 
This particularly convenient basis has been already used in \cite{Dumont:2013wma},
 is discussed more formally in \cite{Einhorn:2013kja}, and is consistent with the general basis presented in \cite{Grzadkowski:2010es}. We display below only the operators which will feed into anomalous gauge couplings, or will be closely related to their study. In particular, dipole operators and Yukawa-like dimension six operators are not shown.

The dimension six Lagrangian $\mathcal L^{(6)}$ contains the operators\footnote{The four fermion operators $\mathcal O_{4f}$ and $\mathcal O_{4f}'$ do not appear in the basis of Ref.~\cite{Dumont:2013wma}, as they can be eliminated in favor of the two operators $(\partial_\rho B_{\mu\nu})^2$ and $(D_\rho W_{\mu\nu}^a)^2$, which played no role in the analysis of that paper.}
\be
\O_{D^2}=|H|^2|D_\mu H|^2\,,\qquad \O'_{D^2}=|H^\dagger D_\mu H|^2\,, \label{HDO_O_6}
\ee
\be
\O_D=J^a_{H\,\mu}\,J^a_{f\,\mu}\,,\qquad
\O_{D}'=J^Y_{H\,\mu}\,J^Y_{f\,\mu}\,,
\ee
\be
\O_{4f}=J^a_{f\,\mu}\,J^a_{f\,\mu}\,,\qquad
\O_{4f}'=J^Y_{f\,\mu}\,J^Y_{f\,\mu}\,,
\ee
\be
\O_{\WW}=H^\dagger H\,(W_{\mu\nu}^a)^2\,,\qquad \label{HDO_O_WW}
\O_{\BB}=H^\dagger H\,(B_{\mu\nu})^2\,,\qquad
\O_{\WB}=H^\dagger\,W_{\mu\nu} H\,B_{\mu\nu}\,,
\ee
\be
\mathcal{O}_{W^3}=\varepsilon^{IJK}W_{\mu\nu}^I W_{\nu\rho}^J W_{\rho\mu}^K\,.
\ee
The SM currents are defined as
\be
J^{a\, \mu}_f=\sum_{\psi_L} \frac{1}{2}\,\bar \psi_L\gamma^\mu \frac{\sigma^a}{2}\psi_L\,\qquad J^{Y\,\mu}_f=\sum_\psi Y_\psi \bar\psi\gamma^\mu\psi
\ee
and
\be
J^{a\,\mu}_H=\frac{i}{2}H^\dagger \sigma^a \overleftrightarrow D^\mu H\,,\qquad 
J^{Y\,\mu}_H=\frac{i}{2}\, H^\dagger\overleftrightarrow D^\mu H
\ee
where $Y_\psi$ is denotes the hypercharge.

The dimension eight Lagrangian relevant for neutral gauge couplings is \cite{Gupta:2011be}
\be
\begin{split}
\mathcal{L}_{(8)}=& \frac{\alpha_1}{\Lambda^4} D^\mu H^\dagger D_\mu H  D^\nu H^\dagger D_\nu H+ \frac{\alpha_2}{\Lambda^4} D^\mu H^\dagger D_\nu H  D^\nu H^\dagger D_\mu H \\
&
+\frac{\alpha_3}{\Lambda^4} D^\rho H^\dagger D_\rho H  B_{\mu\nu}B^{\mu\nu}
+\frac{\alpha_4}{\Lambda^4} D^\rho H^\dagger D_\rho H  W^I_{\mu\nu}W^{I\mu\nu}
+\frac{\alpha_5}{\Lambda^4} D^\rho H^\dagger \sigma^I D_\rho H  B_{\mu\nu}W^{I\mu\nu}\\
&+\frac{\alpha_6}{\Lambda^4} D^\nu H^\dagger D_\rho H  B_{\mu\nu}B^{\mu\rho}
+\frac{\alpha_7}{\Lambda^4} D^\nu H^\dagger D_\rho H  W^I_{\mu\nu}W^{I\mu\rho}\\
&+\frac{\alpha_8}{\Lambda^4} B^{\mu\nu}B_{\mu\nu}B^{\rho\sigma}B_{\rho\sigma}
+\frac{\alpha_9}{\Lambda^4} W^{I\mu\nu}W^I_{\mu\nu}W^{J\rho\sigma}W^J_{\rho\sigma}
+\frac{\alpha_{10}}{\Lambda^4} W^{I\mu\nu}W^J_{\mu\nu}W^{I\rho\sigma}W^J_{\rho\sigma}\\
&+\frac{\alpha_{11}}{\Lambda^4} B^{\mu\nu}B_{\mu\nu}W^{I\rho\sigma}W^I_{\rho\sigma}
+\frac{\alpha_{12}}{\Lambda^4} B^{\mu\nu}W^I_{\mu\nu}B^{\rho\sigma}W^I_{\rho\sigma}\\
&+\frac{\alpha_{13}}{\Lambda^4} B^{\mu\nu}B_{\nu\rho}B^{\rho\sigma}B_{\sigma\mu}
+\frac{\alpha_{14}}{\Lambda^4} W^{I\mu\nu}W^I_{\nu\rho}W^{J\rho\sigma}W^J_{\sigma\mu}
+\frac{\alpha_{15}}{\Lambda^4} W^{I\mu\nu}W^J_{\nu\rho}W^{I\rho\sigma}W^J_{\sigma\mu}\\
&+\frac{\alpha_{16}}{\Lambda^4} B^{\mu\nu}B_{\nu\rho}W^{I\rho\sigma}W^I_{\sigma\mu}
+\frac{\alpha_{17}}{\Lambda^4} B^{\mu\nu}W^I_{\nu\rho}B^{\rho\sigma}W^I_{\sigma\mu}\,.
\end{split}
\ee
We limit ourselves to two-photon QGCs, such that by gauge invariance the operators $\mathcal{O}_{1,2}$ cannot contribute. 


\section{Anomalous gauge couplings in the broken phase}
\label{L_AGC}

\label{anomalous}

In this section we revisit and expand the anomalous Lagrangian in the broken phase, \ie~ expressed in terms of $F_{\mu\nu}$, $Z_\mu$, $W_\mu^\pm$, containing  all leading contributions to anomalous TGCs and QGCs.
 One can distinguish two types of deviations, those correcting couplings already existing in the SM at tree-level, denoted by coefficients $\kappa$, and those corresponding to new Lorentz structures, denoted by coefficients $\lambda$, $\eta$, $\zeta$. The Standard Model corresponds to taking $\kappa_i\rightarrow0$, $\lambda_i\rightarrow0$, $\eta_i\rightarrow0$, $\zeta_i\rightarrow0$.


We organize the Lagrangian as 
\be
\mathcal{L}_{\rm eff}=\mathcal{L}_{\rm kin}+\mathcal{L}_{\rm CGC}+\mathcal{L}_{\rm NGC}\,.
\ee
The first term $\mathcal{L}_{\rm kin}$ contains the kinetic terms and no anomalous interactions. We write it only to set up the conventions.
The second term  $\mathcal{L}_{\rm CGC}$ contains the charged gauge couplings, including the Standard Model ones. The third term  $\mathcal{L}_{\rm NGC}$ contains the neutral gauge couplings, which do not exist in the SM at tree-level.
One can further decompose the anomalous Lagrangians as 
\be
\mathcal{L}_{\rm CGC}=\mathcal{L}_{\rm CGC}^{\textrm{SM},v}+\mathcal{L}_{\rm CGC}^{\partial}\,,\quad
\mathcal{L}_{\rm NGC}=\mathcal{L}_{\rm NGC}^{v}+\mathcal{L}_{\rm NGC}^{\partial}\,,
\ee
where the index $v$ denotes anomalous couplings generated by HDOs involving the Higgs, and thus proportional to $v^2/\Lambda^2$, while the index $\partial$ denotes anomalous couplings generated by pure gauge HDOs, and thus containing higher derivatives like $\partial^2/\Lambda^2$.
We can identify two regimes, the ``broken phase'' regime  $E<v$ in which the effect of the  $\mathcal{L}^\partial$ Lagrangian in the observables can typically be neglected with respect to the effects of $\mathcal{L}^v$ , and the ``unbroken phase'' regime   $v<E<\Lambda$ where the effects of $\mathcal{L}^\partial$ dominate over the ones of $\mathcal{L}^v$. In the regime $\Lambda<E$ the effective field theory prescription breaks down and our results do not apply.

The kinetic Lagrangian reads
\be
\mathcal{L}_{\rm kin}=-\frac{1}{2}\hat W^+_{\mu\nu}\hat W^-_{\mu\nu}
-\frac{1}{4}F_{\mu\nu}^2
-\frac{1}{4}Z_{\mu\nu}^2
\ee
where we defined $U(1)_{em}$ covariant field strength
\be
\hat W^+_{\mu\nu}=D_\mu W^+_{\nu} -D_\nu W^+_\mu\,.
\ee
where here $D$ is only covariant w.r.t.~$U(1)_{em}$. 
The kinetic term $\hat W^+_{\mu\nu}\hat W^-_{\mu\nu}$ contains all $U(1)_{em}$  interactions between the photon and the $W$'s which are protected from any deviations.
The $\mathcal{L}_{\rm CGC}^{\textrm{SM},v}$ piece reads \footnote{The $\kappa$ anomalous couplings are related to 
the customary parametrization of \cite{Hagiwara:1986vm} following $\kappa_\gamma=1+\kappa_1$, $\kappa_Z=1+\kappa_2$, 
$g_1^Z=1+\kappa_3$. The $SU(2)\times U(1)$ gauge invariance implies in general $\kappa_2=\kappa_3- \kappa_1\,\frac{s_w^2}{c^2_w}$.
}
\begin{multline}
\mathcal{L}_{\rm CGC}^{\textrm{SM},v}= -ie (1+\kappa_1)F_{\mu\nu}W_\mu^-W_\nu^+ -ig_Z (1+\kappa_2)Z_{\mu\nu}W_\mu^-W_\nu^+\\
-i g_Z(1+\kappa_3)\left[\hat W_{\mu\nu}^+W_\nu^--\hat W_{\mu\nu}^-W_\nu^+\right] Z_\mu
- g_Z^2(1+\kappa_4)^2\biggl(|W_\mu^+|^2| Z_{\nu}|^2-| Z_{\mu}W_\mu^+|^2\biggr)
\\+\frac{1}{4} g_W^2( 1+\kappa_5)^2( W_\mu^-W_\nu^+-W_\mu^+W_\nu^-)^2
+\eta_1^W F^{\mu\nu}F_{\mu\nu}W^{+\rho}W_{\rho}^-+ 
\eta_2^W F^{\mu\nu}F_{\mu\sigma}W^{+\nu}W_{\sigma}^-
\,,
\end{multline}
where in terms of the operator coefficients 
\be
\begin{split}
\kappa_1&=\frac{\alpha_{WB}}{2t_w}\frac{v^2}{\Lambda^2}\,,\,\,\,\\
\kappa_2&=-\frac{ s_w c_w}{( c_w^2- s_w^2)}\,\alpha_{\WB}\, \frac{v^2}{\Lambda^2}-\frac{1}{4( c_w^2- s_w^2)}[\alpha'_{D^2}+\alpha_D-\alpha_{4f}] \frac{v^2}{\Lambda^2}\,,\\
\kappa_3&=\kappa_4=-\frac{ s_w }{2 c_w( c_w^2- s_w^2)}\,\alpha_{\WB}\,\frac{v^2}{\Lambda^2}-\frac{1}{4( c_w^2- s_w^2)}[\alpha'_{D^2}+\alpha_D-\alpha_{4f}]\frac{v^2}{\Lambda^2}\,,\\
\kappa_5&= -\frac{ c_w s_w}{2( c_w^2- s_w^2)}\alpha_{\WB}\, \frac{v^2}{\Lambda^2}-\frac{ c_w^2}{4(\hat c_w^2- s_w^2)}[\alpha'_{D^2}+\alpha_D-\alpha_{4f}] \frac{v^2}{\Lambda^2}\\
\eta_1^W&=\frac{m_W^2}{\Lambda^4}(c^2_w\, \alpha_3 + s^2_w\, \alpha_4+c_w s_w \,\alpha_5)\\
\eta_2^W&=\frac{m_W^2}{\Lambda^4}(c^2_w\, \alpha_6 + s^2_w\, \alpha_7)
\,.
\end{split}
\ee
The  $\mathcal{L}_{\rm CGC}^{\partial}$ piece reads 
\be
\begin{split}
\mathcal{L}_{\rm CGC}^{\partial}=&
\lambda^Z\left[ig_Z Z_{\mu\nu} (\hat W^-_{\nu\rho}\hat W^+_{\rho\mu}-\hat W^+_{\nu\rho}\hat W^-_{\rho\mu})\right]
+\lambda^\gamma\left[ie F_{\mu\nu}(\hat W^-_{\nu\rho}\hat W^+_{\rho\mu}-\hat W^+_{\nu\rho}\hat W^-_{\rho\mu})\right]\\
&+\zeta^W_1F^{\mu\nu}F_{\mu\nu}W^{+\rho\sigma}W^-_{\rho\sigma}+\zeta^W_2F^{\mu\nu}F_{\nu\rho}W^{+\rho\sigma}W^-_{\sigma\mu}\nn\\
&+\zeta^W_3F^{\mu\nu}W^+_{\mu\nu}F^{\rho\sigma}W^-_{\rho\sigma}+ 
\zeta^W_4F^{\mu\nu}W^+_{\nu\rho}F^{\rho\sigma}W^-_{\sigma\mu} \,.
\end{split}
\ee
 One has 
\be
\begin{split}
\lambda^Z&=\lambda^\gamma=3\frac{\alpha_{W^3}}{g\,\Lambda^2}\\
\zeta_1^W&=\Lambda^{-4}\left(4s^2_w\,\alpha_9+2c^2_w\,\alpha_{11}\right)\\
\zeta_3^W&=\Lambda^{-4}\left(4s^2_w\,\alpha_{10}+2c^2_w\,\alpha_{12}\right)\\
\zeta_2^W&=\Lambda^{-4}\left(4s^2_w\,\alpha_{14}+2c^2_w\,\alpha_{16}\right)\\
\zeta_4^W&=\Lambda^{-4}\left(4s^2_w\,\alpha_{15}+2c^2_w\,\alpha_{17}\right)\,.
\end{split}
\ee

The $\mathcal{L}_{\rm NGC}^{v}$ piece is 
\be
\mathcal{L}_{\rm NGC}^{v}=
\eta^Z_1 F^{\mu\nu}F_{\mu\nu}Z_{\rho}Z_{\rho}+ 
\eta^Z_2 F^{\mu\nu}F_{\mu\sigma}Z_{\nu}Z_{\sigma}\\
\ee
with
\be
\begin{split}
\eta^Z_1 &=\frac{m_Z^2}{2\,\Lambda^4}(c^2_w \,\alpha_3 + s^2_w\, \alpha_4-c_w s_w\, \alpha_5)\\
\eta^Z_2&=\frac{m_Z^2}{2\,\Lambda^4}(c^2_w \,\alpha_6 + s^2_w\, \alpha_7)\,.
\end{split}
\ee

Finally, the $\mathcal{L}_{\rm NGC}^{\partial}$ piece is \footnote{ The neutral coefficients have been computed in \cite{Gupta:2011be}. We observe a discrepancy between our  coefficients $\zeta^\gamma_1$, $\zeta^\gamma_2$
and the corresponding  coefficients $a_1^{\gamma \gamma}$, $a_2^{\gamma \gamma}$ 
 given in \cite{Gupta:2011be}.  }
\begin{align}
\mathcal{L}_{\rm NGC}^{\partial}=
&\zeta^\gamma_1 F^{\mu\nu}F_{\mu\nu}F^{\rho\sigma}F_{\rho\sigma}
+ \zeta^\gamma_2 F^{\mu\nu}F_{\nu\rho}F^{\rho\sigma}F_{\sigma\mu}\nn\\
&+\zeta^{\gamma Z}_1F^{\mu\nu}F_{\mu\nu}F^{\rho\sigma}Z_{\rho\sigma}+ 
\zeta^{\gamma Z}_2 F^{\mu\nu}F_{\nu\rho}F^{\rho\sigma}Z_{\sigma\mu}\nn\\
&+\zeta^Z_1 F^{\mu\nu}F_{\mu\nu}Z^{\rho\sigma}Z_{\rho\sigma}+ 
\zeta^Z_2F^{\mu\nu}F_{\nu\rho}Z^{\rho\sigma}Z_{\sigma\mu}\nn\\
&+\zeta^Z_3F^{\mu\nu}Z_{\mu\nu}F^{\rho\sigma}Z_{\rho\sigma}+ 
\zeta^Z_4 F^{\mu\nu}Z_{\nu\rho}F^{\rho\sigma}Z_{\sigma\mu}\nn\,,
\end{align}
with
\be
\begin{split}
\zeta_1^\gamma&=\Lambda^{-4}\left(c^4_w\, \alpha_8+s_w^4(c_9+\alpha_{10})+c^2_ws^2_w(\alpha_{11}+\alpha_{12})\right)\\
\zeta_1^{\gamma Z}&=\Lambda^{-4}\left(-4 s_w c_w^3 \alpha_{8}+4s_w^3c_w(\alpha_{9}+\alpha_{10})+2c_ws_w(c^2_w-s^2_w)(\alpha_{11}+\alpha_{12})\right)  \\
\zeta_1^Z&=\Lambda^{-4}\left(2c_w^2s^2_w(\alpha_8+\alpha_9+\alpha_{10}-a_{12})+(c^4_w+s^4_w)\alpha_{11}\right) \\
\zeta_3^Z&=\Lambda^{-4}\left(4c_w^2s^2_w(\alpha_8+\alpha_9+\alpha_{10}-\alpha_{11})+(c^2_w-s^2_w)^2\,\alpha_{12}\right)\\
\end{split}
\ee

\be
\begin{split}
\zeta_2^\gamma&=\Lambda^{-4}\left(c^4_w\, \alpha_{13}+s_w^4(\alpha_{14}+\alpha_{15})+c^2_ws^2_w(\alpha_{16}+\alpha_{17})\right)\\
\zeta_2^{\gamma Z}&=\Lambda^{-4}\left(-4 s_w c_w^3 \alpha_{13}+4s_w^3c_w(\alpha_{14}+\alpha_{15})+2c_ws_w(c^2_w-s^2_w)(\alpha_{16}+\alpha_{17}) \right) \\
\zeta_2^Z&=\Lambda^{-4}\left(4c_w^2s^2_w(\alpha_{13}+\alpha_{14}+\alpha_{15}-\alpha_{17})+(c^2_w-s^2_w)^2\alpha_{16} \right)\\
\zeta_4^Z&=\Lambda^{-4}\left(2c_w^2s^2_w(\alpha_{13}+\alpha_{14}+\alpha_{15}-\alpha_{16})+(c^4_w+s^4_w)\,\alpha_{17}\right)\,.
\end{split}
\ee
A summary of our notation for the many anomalous couplings can be found in Tab.~\ref{summary}.
\begin{table}
\begin{center}
\begin{tabular}{|c|cc|cc|}
\hline
	& \multicolumn{2}{c|}{Charged couplings}& \multicolumn{2}{c|}{Neutral  couplings}\\
	& $\mathcal L^{v}_{CGC}$	&$\mathcal L^{\partial}_{CGC}$& $\mathcal L^{v}_{NGC}$	&$\mathcal L^{\partial}_{NGC}$\\
\hline
Triple  couplings (TGCs)	&$\kappa_i$	&$\lambda^{\gamma},\ \lambda^Z$&$-$&$-$\\
Quartic  couplings (QGCs)	&$\eta^W_i$	&$\zeta^W_i$	&$\eta^Z_i$	&$\zeta_i^{\gamma},\ \zeta_i^Z,\ \zeta^{\gamma Z}_i$\\	
\hline
\end{tabular}
\end{center}
\caption{Summary of our notation for the anomalous gauge couplings.}
\label{summary}
\end{table}

Before we close this section, let us comment on the validity of EFT and the experimental sensitivity. 
The operators $\mathcal O_i$ are supposedly generated by a heavy particle of mass $\Lambda$, and hence EFT is valid as long as $ E \ll \Lambda $. 
In practice (i.e.~at a hadron collider) the energy will be distributed over a certain range, but bounded from above by some $E_{\rm max}$, hence it is reasonable to only demand 
\be
 E_{\rm max}\leq\Lambda\,,  
\label{EFTval}
\ee
with the equality still being reasonably consistent with the EFT approach.
Let us define the experimental sensitivity $\Lambda^s_i$ to be the (say) 95\% C.L. bound on the interaction $(\Lambda_i^s)^{-m_i}\mathcal O_i$. Clearly $\Lambda_s$ is a function of $E_{\rm max}$, and a given heavy particle can be constrained only if 
\be
\Lambda\leq (\alpha_i)^{\frac{1}{m_i}}\, \Lambda_s^i(E_{\rm max})\,.
\label{sens}
\ee
One can obtain a rough estimate of the sensitivity by combining Eq.~(\ref{sens})  with the EFT criterion Eq.~(\ref{EFTval}). This implies that 
$
\Lambda_s^i(E_{\rm max})\geq  (\alpha_i)^{-\frac{1}{m_i}}E_{\rm max}
$
in order for the experiment to be sensitive to a particular New Physics effect.
 The ratio $E_{\rm max}/\Lambda_s^i(E_{\rm max})$ is a measure of how strongly coupled new physics has to be in order to be tested experimentally by the effective interaction $\mathcal O_i$.


\section{The generic loop contributions to   $\mathcal{L}^{\partial}$}
\label{Generic_L_partial}

We observed above that the $\mathcal{L}^\partial$ piece of the effective Lagrangian dominates in the $ v<E<\Lambda$ regime, \ie~the  $\mathcal{L}^v$ piece  can be neglected. This implies that the effective Lagrangian is not only $U(1)_{em}$ symmetric, but also $SU(2)_L\times U(1)_Y$ symmetric in this regime. 
Given that $\mathcal{L}^\partial$ contains only pure gauge operators, one can then make the crucial observation that any one-loop diagram contributing to $\mathcal{L}^\partial$ has all its vertices fixed by gauge-invariance.  As a consequence, any perturbative contribution to $\mathcal{L}^\partial$ is fixed once the quantum numbers and the mass of the states running in the loop are specified. This allows us to provide formulas to easily compute $\mathcal{L}^\partial$ in any model of new physics. 
We employ the heat kernel method to  derive these loop contributions. Some elements of the derivations are collected in App. \ref{app_heatkernel}.

Let us consider a (real) heavy vector $X_\mu$ with mass $m_X$ charged under $SU(2)_L\times U(1)_Y\equiv G_{EW}$. 
In order to obtain a consistent Lagrangian, we will write the theory as a nonlinearly realized theory of a larger, not necessarily simple, gauge group $G$, 
i.e. we assume that the heavy gauge bosons live in the coset $G/G_{EW}$.
  The interactions of this heavy vector with the electroweak gauge bosons $V_\mu=(W_\mu,B_\mu)$ are described by
\be
\mathcal{L}=-\frac{1}{4}(\hat X_{\mu\nu}^{\tilde a})^2-\frac{1}{4}(V_{\mu\nu}^a)^2- \frac{c}{2} X_{\mu}^{\tilde a} X_{\nu}^{\tilde b}\,  f^{\tilde{a}\tilde{b}c}\, V^c_{\mu\nu} \,.
\ee
The indices with tildes denote the broken generators. The $f^{ABC}$ structure constants of $G$ split into those of $G_{EW}$, $f^{abc}$, and the coset $f^{\tilde a\tilde b c}$ which form a (in general reducible) representation of $G_{EW}$. 
The  $\hat X_{\mu\nu}^{\hat a}$ is the $G_{EW}$ covariant field strength,
\be
\hat X_{\mu\nu}^{\tilde a}= D_\mu X_\nu^{\tilde a}- D_\nu X_\mu^{\tilde a}
\ee
The coefficient $c$ could in general  be arbitrary, but at leading order is given by $c=g$, if we assume that there is at least a small hiearchy $f\ll\Lambda$ between the cutoff of the theory and the scale $f$ of the breaking $G\to G_{EW}$.
As typical example -- that also serves as a model for the first Kaluza Klein (KK) resonances of an extra dimensional theory -- consider a two-site model with gauge group $G=G'\times G_{EW}$ with $G'\supset G_{EW}$. Then the coset $G/G_{EW}$ contains precisely the adjoint of $G_{EW}$ and the representations appearing in the coset $G'/G_{EW}$, that is, formally all the representations of the decomposition $G'\to G_{EW}$ appear once.

We work in the Feynman-background gauge, in which there is no mixing between spin-1 and Goldstone degrees of freedom, the latter being degenerate in mass with the corresponding gauge fields.
The ghosts contribute as an adjoint with multiplicity $-2$, and they are also degenerate in mass.
Then, we can compute the contribution of each irreducible $G_{EW}$ representation $t^a_X$ appearing in the decomposition under $G\to G_{EW}$ to the one-loop effective action as 
\be
 S^X_{\rm eff}=\frac{i}{2}\tr \log \left(-[D^2+m_X^2]\eta_{\mu\nu}+2ig\, V^a_{\mu\nu}t^a_X\right)+\frac{i}{2}(1-2)\tr\log\left(-D^2-m_X^2\right)
\label{SeffX}
\ee
where the last term is the sum of Goldstone and ghost contributions. Everything is background-covariant w.r.t.~$V_\mu$. One might observe that the massive vector in this gauge is thus a trivial generalization to the massless one, the only differerence besides the nonzero mass being the Goldstone contribution.

Apart from the generic heavy vector, let us also consider a real scalar and a Dirac fermion with masses $m_S$, $m_f$ transforming under respectively $t^a_S$, $t^a_f$ of $G_{EW}$. The corresponding contributions are
\bea
S^S_{\rm eff}&=&+\frac{i}{2}\tr \log \left(-D^2-m_S^2\right)\,,\\
S^f_{\rm eff}&=&-\frac{i}{2}\tr \log \left(-D^2-m_f^2+g\, S^{\mu\nu}\,V^a_{\mu\nu}t^a_f\right)\,,
\label{Sefff}
\eea
where the Lorentz generators are defined as $S^{\mu\nu}=\frac{i}{4}[\gamma^\mu,\gamma^\nu]$.
Denoting the generators of a representation of a particle with spin $s=(0,\frac{1}{2},1)$ as $T^a=\{t^a_S,t^a_f,t^a_X\}$, 
the three and four field--strength operators will be respectively proportional to the traces of three and four generators 
$\tr[T^aT^bT^c]$, $\tr[T^aT^bT^cT^d]$. 
Notice that (irreducible) representations of $G_{EW}$ are simply labelled by the hypercharge $Y$ and by the dimension $d$ of the $SU(2)_L$ representation, which can assume any positive integer value. Therefore, for each $d_Y$, the trace and thus the operator can be computed as a function of $d$ and $Y$ only. 

Here below, although the representation $d_Y$ is in general different for each particle, we will omit the index  ``$S,f,X$''  for simplicity. The evaluation of the one loop effective action is performed in App.~\ref{app_heatkernel}.
The generic three--field strength operators generated by loops of $S,f,X_\mu$ in an arbitrary representation $d_Y$  are
\be  \label{OW3loop}
\mathcal{L}^{(6)}\supset 
\frac{g^3}{16\pi^2}\left(-\frac{1}{144\,m_S^2}+\frac{1}{36\, m_f^2}-\frac{1}{48\, m_X^2}\right) \frac{(d^2-1)d}{24} \mathcal{O}_{W^3}\,, 
\ee
The four--field strength operators are
\be \label{OW4loop1}
\begin{split}
\mathcal{L}^{(8)}\supset&\frac{1}{16\pi^2\,m_S^4}\,\left\{
\frac{1}{576}\mathcal{A}+
\frac{1}{720}\mathcal{B}+\frac{1}{420}\mathcal C+\frac{2}{35}\mathcal D
\right\}\\
&+\frac{1\,}{16\pi^2\,m_f^4}\, \left\{
-\frac{1}{36}\mathcal{A}+
\frac{7}{90}\mathcal{B}
-\frac{64}{105}\mathcal C
+\frac{104}{35}\mathcal D
\right\}\\
&+\frac{1}{16\pi^2\,m_X^4}\, \left\{
-\frac{5}{64}\mathcal{A}+
\frac{27}{80}\mathcal{B}
-\frac{111}{140}\mathcal C+\frac{342}{35}\mathcal D
\right\}\,,
\end{split}
\ee
where we have defined the operator combinations
\be\begin{split} \label{OW4loop2}
\mathcal{A}=&\
g'^4\,d\,Y^4\,\mathcal{O}_{8}+
g^4\,\left( \frac{(d^4-1)d}{240} \mathcal{O}_9+\frac{(d^2-1)(d^2-4)d}{120}\mathcal{O}_{10}  \right)
\\
&+g^2g'^2 \,\frac{(d^2-1)d}{6}Y^2\, (\mathcal{O}_{11}+2
\mathcal{O}_{12})\,,
\end{split} \ee
\be\begin{split}
\mathcal{B}=&\
g'^4\,d\,Y^4\,\mathcal{O}_{13}
+g^4\, \left( \frac{(d^4-1)d}{120}\mathcal{O}_{14} +\frac{(d^2-9)(d^2-1)d}{240} \mathcal{O}_{15} \right)
\\
&+g^2g'^2 \,\frac{(d^2-1)d}{6}Y^2\, (2\mathcal{O}_{16}+
\mathcal{O}_{17})\,, \label{OW4loop3}
\end{split} \ee
\be
\mathcal C=g^4\,\frac{d(d^2-1)}{1152}(\mathcal O_{10}-\mathcal O_9)\,,\qquad
\mathcal D=g^4\,\frac{d(d^2-1)}{1152}(\mathcal O_{15}-\mathcal O_{14})\,,
\ee
A first non-trivial result following from these formulas is that in general the contributions from the scalar are much smaller than the ones from particles of other spin. This is the well known dominance of the "magnetic moment" contributions (the field strength terms present in Eqns.~(\ref{SeffX}) and (\ref{Sefff})) over the "convective" ones (stemming from the covariant derivatives), see e.g.~\cite{Peskin:1995ev}.
For $d=1$, it is clear that only the pure hypercharge contribution remains. 
The operators $\mathcal C$ and $\mathcal D$ do not contribute to NGCs.
One can see that the loop contributions grow respectively as $O(d^3)$ and $O(d^5)$ for the three and four field strength operators. For the latter these are the pure $W$ operators $\mathcal{O}_{9,10,14,15}$ which dominate in this limit.


\section{Composite Higgs models and their holographic description}
\label{CHmodels}

In this section we apply our general formulas to obtain $\mathcal{L}^\partial$ in  composite Higgs models.  In such models  the Higgs is identified as the pseudo Goldstone bosons of an approximate global symmetry $G$, spontaneously broken into $H$ by the dynamics of a new strongly-interacting sector. The strong sector is coupled to a so-called elementary sector containing the SM gauge and fermion fields. The elementary SM fermions get mixed with composite fermions of the strong sector.  This coupling between the two sectors breaks explicitly the strong sector global symmetry, giving a potential to the Higgs, thereby triggering electroweak symmetry breaking. The Higgs sector is fully fixed once the coset $G/H$ is specified. For a given $G$, there is also a freedom to choose the embedding of the composite fermions with which the elementary fermion mixes. 
The top sector plays a central role in this scenario, as a natural realization requires a  composite top partner lighter than the mass scale of the strong sector (see e.g \cite{ratta}).

A composite Higgs model is described by an effective chiral Lagrangian, and contains in general many free continuous parameters. 
However we saw that our generic formulas Eqs. \eqref{OW3loop}--\eqref{OW4loop3} only depend on the quantum numbers and  masses of the resonances, the latter being the only continuous parameters. This is because all relevant couplings are fixed by gauge invariance. 
 This  feature provides a window on the quantum numbers of the model.  Even if a lot of realizations are possible, this still constitutes a discrete set of possibilities, much easier to fit than a set of continuous parameters. A central question is then whether or not one can identify the strong sector global symmetry group through these measurements.


The composite resonances fill  complete multiplets of $G$ denoted $r_G$. To find out how they interact with the SM gauge fields, these multiplets  have to be decomposed under $G\rightarrow G_{EW}$ as $r_G=\bigoplus
d_Y$. 
As the global symmetry is broken, the various components $d_Y$ resulting from a single $r_G$ do not  necessarily have the same mass. Moreove, there is in general a small splitting from electroweak symmetry breaking. However, in our framework we assume $\Lambda\gg v$, so we can neglect the latter.~\footnote{Some of the resonances we discuss below could be rather light. However, their masses have to be somewhat higher than $v$ as they   would already have been detected at the LHC otherwise, see    \cite{deSandes:2008yx,delAguila:2010es,Dissertori:2010ug}.  We can thus safely assume $\Lambda\gg v$ and neglect the electroweak mass splittings.}
   The picture of the mass spectrum is therefore as follows. The various $r_H$ components of  $r_G$  $(r_G=\bigoplus r_H)$ can have in general different masses. In addition,  the $d_Y$ components mixing with the SM fermions (mostly in the top sector) get a different mass from the rest of the $r_H$ multiplet. Otherwise, components of $r_H$  are all  degenerate. 

Let us focus on the  global symmetry group $G\equiv SO(4+N)$, used in a large part of composite Higgs models. In practice, $N$ ranges from $1$ to $5$. 
 We do not discuss  contributions from possible scalar resonances as  they are negligible with respect to the ones from fermion and vector resonances.
 
The gauge resonances of these models decompose under $G\rightarrow G_{EW}$ as
\be 
\mathbf{Adj}(G)\rightarrow \mathbf{3}_0\oplus\mathbf{1}_{-1}\oplus\mathbf{1}_{1} \oplus N\times \left(\mathbf{2}_{1/2}\oplus\mathbf{2}_{-1/2}  \right)\oplus  \left(\frac{N(N-1)}{2}+1\right)\times\mathbf{1}_{0}
\,.\ee
The $\mathbf{3}_0$ and $\mathbf{1}_0$ are SM-like resonances corresponding to heavy copies of the $W^I_\mu$, $B_\mu$ fields. We can see that a  number of degrees of freedom of $\mathbf{Adj}(G)$ goes into the singlets, which do not contribute to the anomalous gauge couplings. Amongst the non SM-like resonances there are $N$ doublets, as well as the (complex) $SU(2)$ singlet of unit hypercharge that is needed to complete $SU(2)_R$ multiplet. Computing the effect from these additional resonances in Eqs.~\eqref{OW3loop}--\eqref{OW4loop3}, we find they typically modify the contributions from the SM-like resonance by $O(10)$ factors. 
We do not discuss further these vectorial resonances as they are in general overwhelmed by fermionic ones.

Considering fermion resonances, we focus our attention on light composite top partners $\Psi$.\footnote{Other light vector-like fermions, such as $\tau$ or bottom partners, can be present in some models. Our results trivially generalize to these cases.} Because they are typically lighter than other states of the strong sector, their contribution may naturally become the dominant one because of the quadratic and quartic mass dependence. 
This state may in general have other light companions coming from the embedding in $r_G$. To stay general, let us consider all components of $r_G$, with possibly different masses.
Getting the correct SM fermions hypercharge requires  to add a $U(1)_X$ charge to $G$,  and that the  quantum numbers of $\Psi$ under $SO(4)\times U(1)_X$ be  $\mathbf{1}_{2/3}$ or $\mathbf{4}_{2/3}$.\footnote{Notice $\mathbf{4}_{2/3}$ contains both top-like  and exotic components.} 
  We consider two typical embeddings of the top partner, in the fundamental $\mathbf{F}_{2/3}$  and in the symmetric traceless $\mathbf{S}_{2/3}$ representations of $G$. These representations decompose respectively under the $d_Y$ components of $G_{EW}$ as
\be\begin{split}
\mathbf{F}_{2/3}&=\mathbf{2}_{1/6}\oplus\mathbf{2}_{7/6}\oplus N\times\mathbf{1}_{2/3}\,,\\
\mathbf{S}_{2/3}&=\mathbf{3}_{5/3}\oplus\mathbf{3}_{2/3}\oplus\mathbf{3}_{-1/3}\oplus N\times\left(\mathbf{2}_{1/6}\oplus+\mathbf{2}_{7/6} \right)\oplus\frac{N(N+1)}{2}\times \mathbf{1}_{2/3}\,.
\end{split}
\ee
The $\mathbf{2}_{1/6}$, $\mathbf{1}_{1/3}$ are top-like components, potentially mixing with the elementary top.
 Because of the $U(1)_X$ charge, all states including singlets are charged under $G_{EW}$.
The $\mathbf{F}_{2/3}$ contains an exotic doublet with high hypercharge and a number of singlets. 
For the $\mathbf{S}_{2/3}$, we see there are several top-like resonances, and new  exotic partners charged in the $SU(2)_L$ adjoint. 
 The pattern of anomalous gauge couplings generated by the  light top partners  is shown in  Tab.~\ref{table_lighttop}. 
 
 We see that the exotic contributions dominate over those of the SM-like resonances because of their quantum numbers. Therefore, unless the exotic resonances are heavier than the SM-like ones, they are dominant in  the anomalous gauge couplings.  
These exotics resonances may therefore be the first manifestation of scenario with strongly interacting Higgs. 
If anomalous gauge couplings are indeed measured, the next step would be to identify the pattern of deviations. The first natural analysis would be to test the hypothesis of a single dominant resonance, comparing the data to the various patterns shown in  Tab. \ref{table_lighttop}. 
On the other hand, because of the possible mass splittings, identifying the global symmmetry group using the mild dependence on $N$  would probably be very challenging.


\begin{table} 
\begin{center}
\begin{tabular}{|c|c|c|c|c|}\hline
& $\mathbf{1}_{2/3}$ & $\mathbf{2}_{1/6}$  & $\mathbf{2}_{7/6}$ & $\mathbf{3}_{5/3}\oplus \mathbf{3}_{2/3}\oplus \mathbf{3}_{-1/3}$ \\
\hline
$\lambda^{Z,\gamma}$ 	& $0$ 	& $ 1.7\times 10^{-4}$	 & $1.7\times 10^{-4}$		& $0.002 $   \\
$\zeta_1^\gamma$ 		& $ -1.0\times 10^{-6}$  	& $-1.1 \times 10^{-6}$ & $-4.1 \times 10^{-5}$  &  $-3.6 \times 10^{-4}$ \\
 $\zeta_1^{\gamma Z}$    &  $ 2.2\times 10^{-6}$ & $ -5.8\times 10^{-6}$ & $-1.7\times 10^{-5}$ & $-4.9\times 10^{-4}$ \\
$\zeta_1^Z$ & $ -0.6 \times 10^{-6}$  & $-4.3\times 10^{-6}$ & $-1.3 \times 10^{-5}$ & $-2.6 \times 10^{-4}$ \\
$\zeta_2^\gamma$ &$ 2.9\times 10^{-6}$ & $1.2 \times 10^{-6}$ & $11 \times 10^{-5}$ & $9.7 \times 10^{-4}$ \\
 $\zeta_2^{\gamma Z}$  & $ -6.2\times 10^{-6}$ & $3.0\times 10^{-6}$ & $3.4\times 10^{-5}$ & $11\times 10^{-4}$ \\
$\zeta_2^Z$ & $ 3.4 \times 10^{-6}$&  $-0.3 \times 10^{-6}$& $5.0 \times 10^{-5}$ &  $8.6 \times 10^{-4}$ \\
$\zeta_3^Z$ &$ -1.2 \times 10^{-6}$ & $-8.6 \times 10^{-6}$ & $-2.7 \times 10^{-5}$ & $-5.1 \times 10^{-4}$ \\
$\zeta_4^Z$ & $ 1.7 \times 10^{-6}$ &  $-0.1\times 10^{-6}$& $2.5 \times 10^{-5}$ & $4.3 \times 10^{-4}$ \\
$\zeta_1^W$ & $0$ & $-2.2 \times 10^{-6}$ & $-6.2\times 10^{-5}$ &  $-7.4 \times 10^{-4}$ \\
$\zeta_2^W$ & $0$ & $-12 \times 10^{-6}$ & $32 \times 10^{-5}$ & $35 \times 10^{-4}$ \\
$\zeta_3^W$& $0$ & $-13 \times 10^{-6}$ & $-13\times 10^{-5}$ & $ -16\times 10^{-4}$\\ 
$\zeta_4^W$& $0$ & $ 22\times 10^{-6}$ & $19 \times 10^{-5}$ & $ 23\times 10^{-4}$\\ 
\hline
\end{tabular} 
\end{center}
\caption{ \label{table_lighttop}  Pattern of anomalous couplings generated by the various top partners from  $\mathbf{F}_{2/3}$ and $\mathbf{S}_{2/3}$ in units of $\Lambda=m_{\Psi}$. The masses of resonances on each column are potentially different. }
\end{table}

Overall the generic perturbative contributions described above display interesting perspectives  for the search and study of composite Higgs models in TeV precision physics, although the predicted deviations are small and rather high experimental precision is needed. 
 However, although these contributions to the $\mathcal{L}^\partial$ Lagrangian are well under control, they might not be the only ones present. In the paradigm of a conformal strong sector  spontaneouly broken in the IR, the theory also contains a light scalar, pseudo Goldstone boson of the broken scale invariance, the so-called dilaton. The couplings of the dilaton  to the rest of particles are strongly constrained by the nonlinearly realized conformal symmetry. It couples to the trace of the energy-momentum (EM) tensor of the elementary and strong sectors. The trace of the EM tensor of the massless (elementary) gauge fields vanishes classically but not at the quantum level. The dilaton couples to the large trace anomaly generated by the running of the gauge coupling driven by composite states above the scale of conformal breaking, see e.g \cite{Chacko:2012vm} and references therein. \footnote{There are also small contributions to trace anomaly from loops of elementary fields and from a possible source of explicit CFT breaking.}
The tree-level exchange of a dilaton would thus generate operators of the form \be 
F^{\mu\nu}F_{\mu\nu}F^{\rho\sigma}F_{\rho\sigma}\,,\quad F^{\mu\nu}F_{\mu\nu}Z^{\rho\sigma}Z_{\rho\sigma}\,,\quad
F^{\mu\nu}F_{\mu\nu}W^{+\,\rho\sigma}W_{\rho\sigma}^-\,
\ee
contributing to the $\mathcal{L}^\partial$  Lagrangian. The Lorentz structure implies that only the $\zeta_1$ couplings  can be modified by the dilaton, while the others remain unaltered.

To get a handle on the leading contributions from the dilaton, let us take the crucial assumption that the strongly interacting CFT has a large number of colors $N_c$. In that case one can invoke the qualitative version of the $AdS/$CFT correspondence, to describe the composite Higgs models in terms of a weakly-coupled five-dimensional theory.  The KK modes of this warped extra dimension are identified as the resonances of the 4d strong sector, while the dilaton of the 4d theory is identified with the radion mode.
However, in addition to describing the dilaton, a striking phenomenological prediction of the $AdS_5$ picture is the presence of light KK gravitons near the IR brane. This implies in the 4d picture the existence of a set of composite spin-$2$ resonances coupling to the EM tensor of the other fields present.

We are now going to switch to the 5d picture, to compute all these effects together.
Following our line of work , we will try to  provide results as model-independently as possible.

\section{Integrating out a  warped extra-dimension}
\label{ADS5}

We now turn to  5d theories with a warped background. We  define an effective framework including the Randall-Sundrum models as well as the pNGB composite Higgs models of Sec.~\ref{CHmodels}.
The framework is described below, and details of the calculations are collected in the Appendix. 
 Let us briefly summarize the configuration of the effective framework. 
\begin{itemize}
\item We describe the 5d background by two free parameters $\tilde k$ and $\kappa$, that are related to the  curvature $k$ and the warp factor $\epsilon$ as $\kappa=k/M_{Pl}$ and $\tilde k=\epsilon k$ respectively.
\item The Higgs localization is controled by the parameter $\nu$, where $\nu=\infty$ is an IR brane Higgs and $\nu=0$ corresponds to a composite pNGB Higgs.
\item The radion has mass $m_{\phi}$, while its couplings are fixed by $\kappa, \tilde k$.
\item Gauge fields are either on the IR brane or in the bulk. If they are in the bulk, the gauge group is either the SM or extended in the custodial and/or gauge-Higgs unification models.
\item In the bulk gauge fields case, $SU(2)_L$ and $U(1)_Y$ IR brane  kinetic terms (BKTs) respectively parametrized as $r$, $r'$  are present.
\end{itemize}


\subsection{The 5d framework}

\label{sec:5d}

The spacetime metric of theories with a warped extra dimension can be written in general as
\be
ds^2= \gamma_{MN}\,dx^Ndx^M= a(z)^{-2}(\eta_{\mu\nu}dx^\mu dx^\nu - dz^2)\,,
\label{metric5d}
\ee
with $\eta_{\mu\nu}=diag(1,-1,-1,-1)$, $z\in[z_{UV},z_{IR}]$.
The  action reads 
\be
\begin{split}
S=-\int dx^5 \sqrt{-g} \{ &M^3(\mathcal{R}+\Lambda)+ \frac{1}{4}F_{MN} F^{MN}-D_M H^\dagger D^M H-\frac{i}{2}\bar\Psi\Gamma^M \overleftrightarrow D_M\Psi \\
+& m^2_H| H|^2+m_\Psi\bar\Psi\Psi+ \mathcal{B}
  \}\,.
  \end{split}\label{L5d}\ee
Here  $M$ is the 5d Planck mass, $\mathcal{R}$ is the Ricci scalar and $\Lambda$ is a negative cosmological constant.   $\mathcal{B}$ contains the  boundary terms, and
$m^2_H$ and $m_\Psi$ denote the 5d masses responsible for the localization of the Higgs and fermion zero modes.
For the purpose of deriving the low-energy effective Lagrangian, there is in general no need to specify a particular background, many analytic expressions can be derived  in terms of $a(z)$ and zero mode profiles \cite{Cabrer:2011fb}. 
Any stabilization mechanism will induce such a nontrivial metric, we will work under the approximation that such backreaction is neglected.

The $AdS_5$ metric $a(z)=kz$ then follows from Eq.~(\ref{L5d}) with $\Lambda=12 k^2$. 
The quantity $k$ is the inverse curvature radius, we take the location of the UV boundary to be $z_0=1/k$ without loss of generality and define $\tilde k\equiv 1/z_{1}$ where $z_1$ is the location of the IR boundary. 
The 4d reduced Planck mass is related to the scale $M$ as $M^3=kM_{Pl}^2$.
The $AdS_5$ background can be described in terms of two parameters, for our purpose it is appropriate to choose the IR scale $\tilde k$ and the dimensionless quantity 
\be \kappa=k/M_{Pl}\,.\ee
The 5d theory is perturbative for values of $\kappa$ of $O(1)$ or smaller. This is obtained by requiring higher curvature terms $\mathcal{R}_5/M_*^2$ to be smaller than $1$. Here $\mathcal{R}_5=20k^2$ is the size of the $AdS$ curvature and $M_*$ is seen as the cutoff of the 5d theory, estimated by naive dimensional analysis \cite{Chacko:1999hg, Agashe:2007zd},  to be $M_*^3\approx (24\pi^3) M_{Pl}^2 k $. Note  that the narrow-width approximation breaks down for $\kappa\gtrsim 2$ $(\kappa\gtrsim 0.3)$
for the case of bulk (IR brane) localized SM fields \cite{Davoudiasl:1999jd,Agashe:2007zd}.
This will not constitute a problem for integrating out KK gravitons, as for this purpose one goes in the zero momentum limit for which the imaginary part of the self-energy vanishes.

Finally, we include the IR brane kinetic terms
\be
\mathcal{B}\supset \delta(z-z_1)\frac{r_I}{4}F^I_{\mu\nu}F^{I\,\mu\nu}\,.
\ee
These  BKTs are generically present in the theory, and are radiatively generated even if they are set to zero at a given scale. These BKTs induce various effects which turn out to be crucial for the phenomenology of the bulk gauge scenario, in particular they distort the gauge KK spectrum. 
The physical gauge coupling is given by \footnote{ We restrict to IR BKTs. In presence of UV BKTs, the physical gauge coupling becomes
$
(g^I)^2=(g^I_5)^2/(V+r_0^I+r_1^I)\,.$
 }
\be
(g^I)^2=\frac{(g^I_5)^2}{V+r_1^I}\,.
\label{gcphys}
\ee
Equation (\ref{gcphys})  constrains $r^I>-V$.

The first KK modes have a mass related to the IR scale $\tilde k$ by $\mathcal{O}(1)$ factors. 
The first graviton and gauge modes without BKTs have  a mass
\be
 m_{\mathbf{2}}\approx 3.8 \,\tilde k \,,\quad m_{\mathbf{1}}\approx 2.4 \,\tilde k\,.
\ee
The first KK gauge modes in presence of respectively positive and negative BKTs $|r_I|> O(1)$ have a mass \footnote{For $r_I\rightarrow+\infty$ the mass matches a Neumann-Dirichlet boundary condition.}
\be
m_{\mathbf{1}}\approx \sqrt{2}\sqrt{\frac{r_I+V}{r_I\, V}} \,\tilde k\,,\quad m_{\mathbf{1}}\approx 3.9\,\tilde{k} \,.\label{eq:gauge_light_mode}
\ee
The first KK fermions  with flat profile have  a mass $m_{\ferm}\approx 2.4 \,\tilde k $.
However, fermions with mixed Neumann-Dirichlet boundary conditions and the appropriate sign of bulk mass can become exponentially light with respect to $\tilde k$. It is important to keep this possibility in mind as it can have consequences for the size of the anomalous gauge couplings. This constitutes the 5d version of the light top partners of the composite Higgs setup.
The other quantities commonly used in the $AdS_5$ background are the warp factor $\epsilon\equiv\tilde k/k$ and the volume factor $V=\log k/\tilde k $. 
Notice that in order to solve the hierarchy problem, i.e. $\tilde k\approx 1~$TeV, one needs $V\approx 37$.

Defining $\nu=\sqrt{m_H^2/k^2+4}$, the Higgs profile along the fifth dimension is 
\be
f_h(z)\propto z^{2+\nu}\,.
\ee
In the limit $\nu\rightarrow \infty$, the wave-function is totally localized on the IR brane, \ie~it can be described by a boundary 4d Lagrangian $\mathcal{L}\supset\delta(z-z_{1})\,\mathcal{L}^{4d}_{H} $.
The lower bounds is $\nu= 0$, \footnote{This corresponds to $m_H^2=-4k^2$, the Breitenlohner-Freedman bound for stability of $AdS$ space.}
for this value the Higgs field is in the bulk, but is still localized towards the IR brane. 

The case $\nu=0$ also describes the couplings of a possible zero mode of the fifth component of a gauge field, $A_5$.  Other properties of a $A_5$ like the 5d energy-momentum tensor are different from those of a fundamental scalar. However, for our purpose, only the tree-level profile matters. We can therefore interpret our $\nu=0$ fundamental Higgs as an $A_5$. 
Identifying the Higgs as a $A_5$ is the central idea of gauge-Higgs unification models.
In the holographic picture, it corresponds to the Goldstone boson of the global symmetry spontaneously broken by the strong dynamics. The $\nu=0$ case corresponds therefore to a holographic description of the composite pNGB models that we discussed in Sec.~\ref{CHmodels}.

The radion in a slice of pure $AdS$ is massless. In the holographic picture it corresponds to the dilaton, i.e.~the Goldstone boson of scale invariance spontaneously broken by the strong dynamics.  Once the size of the extra-dimension is fixed by a  stabilization mechanism, the $AdS$ background gets deformed and the radion obtains a mass. In the holographic picture this corresponds to an explicit deformation of the CFT by a relevant operator. 
In many classes of models the radion is lighter than the IR scale, $m_{\phi}\ll \tilde k$. In this limit its profile stays unperturbed to leading order, its couplings are given by those of a massless radion up to corrections of $\mathcal O(m^2_\phi/\tilde k^2)$,
and the only free parameter is its mass $m_{\phi}$. 

Contrary to scalar fields, gauge fields cannot be continuously localized from bulk to brane. We consider two separate cases, with gauge fields propagating in the bulk and on the IR brane respectively. 
The case of bulk gauge fields corresponds to a global symmetry of the CFT, and one can extend the bulk gauge group to a larger  symmetry including in particular the custodial $SU(2)_R$, as motivated by electroweak precision observables (see Sec.~\ref{constraints}).  We will consider both the non-custodial and custodial cases.

Hereafter we derive the effective 4d Lagrangian arising from this 5d framework.   
We are going to integrate out the radion and KK excitations of gauge, fermions, and gravity. The effective 4d action resulting from integrating out gravity in a slice of $AdS_5$ has been computed in Ref.~\cite{Dudas:2012mv} and is reviewed in App.~\ref{app_WED}. The SM  fields couple to 5d gravity via the modified 5d energy-momentum tensor $\bar T_{MN}$ defined in Eq.~(\ref{mod}). 
The piece proportional to $(\bar T_{\mu\nu})^2$ reads
%
\be
 \mathcal L_{\rm eff}=\frac{1}{4\,M^3}\int_{z_0}^{z_1} dz\ (kz)^3\,
    \left[\Theta_{\mu\nu}(z)-\Omega_{-2}(z)\,\Theta_{\mu\nu}(z_1)\right]^2
\ee
where $\Theta_{\mu\nu}$ is the $\bar T_{\mu\nu}$ integrated over the 5th dimension (see Eq.~(\ref{integrated})) and $\Omega_p$ is defined as 
\be
\Omega_p=\frac{z^p-z_0^p}{z_1^p-z_0^p}\,.
\ee
The leading contribution proportional to $(\bar T_{55})^2$ comes from the radion. Its Lagrangian reads
\be
\mathcal L_{\phi}=\frac{1}{2}(\partial_\mu \phi)^2-\frac{1}{2}\, m_\phi^2\phi^2+\sqrt{\frac{2}{3}}\, \frac{\epsilon}{M_{Pl}}\,\Theta(x,z_1)\, \phi
\ee
where the source $\Theta$ equals $\bar T_{55}$ integrated over the extra dimensional coordinate, see Eq.~(\ref{sources}) for an explicit expression.
Integrating out $\phi$ leads to
\be
\mathcal L_{\rm eff}=\frac{\epsilon^2}{3 M_{Pl}^2}\ \frac{1}{m_\phi^2}\ \left[\Theta(x,z_1)\right]^2\,.
\ee
%

 Integrating out the tree-level KK gauge fields leads to an effective action
\be
\mathcal L_{\rm eff}=
-\frac{(g_5^I)^2}{2}\ a^I_{XY}\ J^{I\,\mu}_X\ J^I_{Y\,\mu}
\ee
where $X,Y=H,f$ label the Higgs and fermions zero modes and $I$ labels the 5d gauge field. The $J_X^{I\,\mu}$ are the zero mode currents.
The quantities $a_{XY}$ are given by 
\be
a_{XY}^I=\int dz\ kz\,\left(\Omega_X-\Omega_I\right)\left(\Omega_Y-\Omega_I\right)
\ee
where \footnote{In presence of both UV an IR BKTs, $\Omega_I=\frac{\log(kz)+r_0^I}{V+r_0^I+r_1^I} $ for ++ gauge fields. }
\be
\Omega_I=\left\{
\begin{array}{cl}
\frac{\log(kz)}{V+r_1^I}&(+\!+{\rm \ gauge\ field)}\\
0&(+\!-{\rm \ gauge\ field)}\\
1&(-\!+{\rm \ gauge\ field)}
\end{array}
\right.
\ee
%
 We are only interested in the oblique and flavor diagonal part of the effective action and therefore can set universally $\Omega_f=1$ as if all fermions were exactly UV localized.\footnote{Deviations from UV localization lead to non-oblique operators for heavy fermions, such as anomalous $Zbb$ couplings that we do not discuss in this paper.}
The integrated profile of the Higgs is given by $\Omega_H=\Omega_{2(\nu+1)}$.
Details of the calculations are left for App.~\ref{app_WED} and can be found in \cite{Cabrer:2011fb,Dudas:2012mv}.

Finally, the loops of KK modes are already included in the general formulas of Sec.~\ref{Generic_L_partial}. The only work to do is to sum the contributions over the tower of KK modes. This translates to small enhancement factors, as the operators we consider are UV-finite, and thus dominated by the lightest KK modes. 
The enhancement factors for the inverse-quartic sum are rather close to one. The zeroes of the Bessel functions $J_\rho$ satisfy~\cite{Elizalde:1992jb} 
\be
\sum_{n} \frac{1}{(m_{\rho,n})^4}=\frac{1}{(2\tilde k)^4(\rho+1)^2(\rho+2)}
\ee 
For instance, for a KK gauge mode with $J_0(m_{\bf 1}/\tilde k)=0$ one has to make the replacement
\be
\frac{1}{(2.405)^4}\approx\frac{1}{33.5}\to \frac{1}{32}
\ee
i.e., a correction factor $K^{(4)}_{\rho= 0}\approx 1.045$.
The correction increases with $\rho$, e.g.~$K^{(4)}_{\rho=1}\approx 1.123$.
The correction are larger for the inverse-square sum, entering in the computation of $\mathcal O_{W^3}$,
\be
\sum_{n} \frac{1}{(m_{\rho,n})^2}=\frac{1}{(2\tilde k)^2(\rho+1)}
\ee 
e.g.~$K^{(2)}_{\rho=0}=1.45$ and $K^{(2)}_{\rho=1}=1.83$.

\subsection{Contributions from KK gravity }

Here we derive the contributions to the effective Lagrangian from KK-gravitons and from the radion. 
We distinguish two separate cases depending whether gauge fields are confined on the IR brane or propagate in the bulk. 
For simplicity, a common BKT has been assumed, $r_I=r$. 
Note in the  numerical results in Secs.~\ref{constraints} and \ref{results} we keep the dependence on both $r$ and $r'$.

For gauge fields in the bulk, the KK gravitons contribute as 
\be\begin{split}
\mathcal{L}^{(8)}\supset& \frac{\kappa^2}{128\,\tilde k^4}\ \frac{1+4r+8r^2}{(r+V)^2}\left(
-\frac{1}{4}( \mathcal{O}_{8}+\mathcal{O}_{9}+2\mathcal{O}_{11})+ \mathcal{O}_{13}+\mathcal{O}_{14}+2\mathcal{O}_{16} 
\right) \\
&+\frac{\kappa^2}{64\,\tilde k^4}\ \frac{1}{r+V}\ \frac{(1+\nu)(5+\nu+4r(3+\nu))}{(3+\nu)^2}\biggr(
-4\mathcal{O}_3-4\mathcal{O}_4  +\mathcal{O}_6+\mathcal{O}_7
\biggr)\,.
\end{split}
\label{L8grav}
\ee
Integrating out the radion leads to
\be 
\mathcal{L}^{(8)}\supset\frac{\kappa^2}{192\,\tilde k^2}\ \frac{1}{(V+r)^2}\ \frac{1}{m_\phi^2}
\left( \mathcal{O}_{8}+\mathcal{O}_{9}+2\mathcal{O}_{11}\right)\,.\label{L8rad}
\ee
The radion also contribute to operators of $\mathcal{L}^{(6)}$,
\be
\mathcal{L}^{(6)}\supset\frac{\kappa^2}{12\,\tilde k^2}\ \frac{1}{V+r}\ \frac{m_h^2}{m_\phi^2}\biggr( \mathcal{O}_{WW}+\mathcal{O}_{BB} +\mathcal{O}_{GG}   \biggr)\label{L6rad}
\,,
\ee
where $m_h$ is the physical Higgs mass.
This contribution can be see as a manifestation of Higgs-radion mixing in the range  $m_h<m_\phi<\tilde k$. 
KK gravitons do not contribute to these operators as their contribution is proportional to the trace of the gauge EM tensor and thus exactly zero.

We finally remark that there is a contribution to the operator $\mathcal O_D^2$ from 5d gravity. This has been computed in a model of 5d supergravity coupled to matter fields in Ref.~\cite{Dudas:2012mv}, here we simply take the result from there by omitting the contributions from superpartners. One finds
\be
\mathcal L^{(6)}\supset -\frac{ \kappa^2}{6\, \tilde k^2}\ \frac{(1+\nu)^2}{3+2\nu}\ \mathcal O_{D^2}
\ee
As already noticed in Ref.~\cite{Dudas:2012mv}, this result diverges in the localization limit $\nu\to \infty$. The reason for this is that 5d gravity couples to 5d mass terms, and hence perturbation theory breaks down for large (but in principle finite) $\nu$. However, starting directly from brane-localized fields (without taking the limit $\nu\to\infty$) one can show that these contributions are exactly zero. 

For gauge and matter fields on the IR brane, one can simply take the limit $r,\nu\to\infty$ in Eqns.~(\ref{L8grav}), (\ref{L8rad}) and (\ref{L6rad}), leading to
\be\begin{split}
\mathcal{L}^{(8)}\supset &\frac{\kappa^2}{16\,\tilde k^4}\left(
-\frac{1}{4}( \mathcal{O}_{8}+\mathcal{O}_{9}+2\mathcal{O}_{11})+ \mathcal{O}_{13}+\mathcal{O}_{14}+2\mathcal{O}_{16} 
\right)\\
&+\frac{\kappa^2}{16\,\tilde k^4}
\biggr(
-4(\mathcal{O}_3+\mathcal{O}_4)  +\mathcal{O}_6+\mathcal{O}_7
\biggr)\,.
\end{split}
\ee
In case of brane localized fields, the radion couples only to the trace of the EM tensor, whose gauge part vanishes. It therefore does not contribute to any operators involving gauge fields.

\subsection{Contributions from  KK gauge modes }

Let us define the functions of the Higgs bulk mass
\be
f_1(\nu)= \frac{2(1+\nu)^2}{(2\nu+3)(\nu+2)}\,,\quad
f_2(\nu)=\frac{(1+\nu)(3+\nu)}{(2+\nu)^2}\,.
\ee
For a brane Higgs, their values are $f_1(\infty)=f_2(\infty)=1$. For a pseudo-Goldstone Higgs, $f_1(0)=1/3$, $f_2(0)=3/4$.  We do not display explicitely the $r$-dependent terms, nor contributions that are subleading in $1/V$. Both can be easily obtained from our general expressions.
In the non-custodial case, one obtains
\be
\begin{split}
\mathcal{L}^{(6)}\supset& -\frac{ 3g^2+g'^2}{8\tilde k ^2} V f_1(\nu) \O_{D^2}+
\frac{ g^2}{4\tilde k ^2} f_2(\nu) \mathcal{O}_D-\frac{ g^2}{\,\tilde k ^2} \frac{1}{8V}  \O_{4f}\\
& -\frac{ g'^2}{4\tilde k ^2} V f_1(\nu) \O_{D^2}'
+\frac{ g'^2}{4\tilde k ^2} f_2(\nu)\mathcal{O}_D'-\frac{ g'^2}{\,\tilde k ^2} \frac{1}{8V}  \O_{4f}'\,.
\end{split}
\ee
In the custodial case, \footnote{The custodial  $\mathcal{O}_{D^2}$ also gets a subleading coefficient $
-\frac{ g'^2}{8\tilde k ^2} f_2(\nu) $.  } we get

\be
\begin{split}
\mathcal{L}^{(6)}\supset& -\frac{ 3g^2+3g'^2}{8\tilde k ^2} V f_1(\nu) \O_{D^2}
+\frac{ g^2}{4\tilde k ^2} f_2(\nu)\mathcal{O}_D-\frac{ g^2}{\,\tilde k ^2} \frac{1}{8V}  \O_{4f}\\
&  +\frac{ g'^2}{4\tilde k ^2} f_2(\nu)\mathcal{O}_{D^2}'
+\frac{ g'^2}{4\tilde k ^2} f_2(\nu)\mathcal{O}_D'-\frac{ g'^2}{\,\tilde k ^2} \frac{1}{8V}  \O_{4f}'\,.
\end{split}
\ee

\subsection{Contributions from KK-loops }

From the KK Higgs, we find\footnote{This contribution only applies to bulk scalars and not to the case of a pGB Higgs.}
\be
 \mathcal{L}^{(6)}\supset -\frac{g^2}{16\pi^2}\frac{1}{288}\frac{K^{(2)}_{\mathbf{0}}}{m_{\mathbf{0}}^2}\,\mathcal{O}_{W^3}\,,
\ee
\be
\begin{split}
\mathcal{L}^{(8)}\supset &\frac{g^4}{16\pi^2}\frac{K^{(4)}_{\mathbf 0}}{4\,m^4_\mathbf{0}}\left(
\frac{33 \mathcal{O}_9+2\mathcal{O}_{10}+8\mathcal{O}_{14}+20\mathcal{O}_{15}  }{20160}
\right)\\
&+\frac{g^2g'^2}{16\pi^2}\frac{K^{(4)}_{\mathbf 0}}{4\,m^4_\mathbf{0}} \left(\frac{1}{576}(2\mathcal{O}_{11}+4\mathcal{O}_{12})+\frac{1}{720}(4\mathcal{O}_{16}+2\mathcal{O}_{17})\right)
\\
&+\frac{g'^4}{16\pi^2}\frac{K^{(4)}_{\mathbf 0}}{4\,m^4_\mathbf{0}}\left(\frac{1}{576}\mathcal{O}_8+\frac{1}{720}\mathcal{O}_{13}\right)
\,,
\end{split}
\ee
where $m_{\bf 0}$ is the lightest KK Higgs mass and the correction from the KK tower amounts to a very mild enhancement factor $K^{(4)}_{\mathbf 0}$. As already stated, contributions of spin-zero states are suppressed compared to those of nonzero spin.  The Higgs KK tower thus plays no role and we will neglect it in what follows.
From the KK SM-like fermions, we find 
\be
 \mathcal{L}^{(6)}\supset \frac{g^2}{16\pi^2}\frac{1}{12}\frac{K^{(2)}_{\nicefrac{\mathbf{1}}{\mathbf{2}}}}{m_{\nicefrac{\mathbf{1}}{\mathbf{2}}}^2}\,\mathcal{O}_{W^3}\,,
\ee
as well as
\be
\begin{split}
 \mathcal{L}^{(8)}\supset& \frac{g^4}{16\pi^2}\frac{K^{(4)}_{\nicefrac{\mathbf{1}}{\mathbf{2}}}}{\,m_{\nicefrac{\mathbf{1}}{\mathbf{2}}}^4} 
 \left( \frac{- 3 \mathcal{O}_{9}-32 \mathcal{O}_{10} + 40 \mathcal{O}_{14} + 58 \mathcal{O}_{15} }{840}\right)
  %
%
\\
&+\frac{g^2g'^2}{16\pi^2}\frac{K^{(4)}_{\nicefrac{\mathbf{1}}{\mathbf{2}}}}{\,m_{\nicefrac{\mathbf{1}}{\mathbf{2}}}^4}\left(
-\frac{1}{36}(\mathcal{O}_{11}+2\mathcal{O}_{12}) +\frac{7}{90}(2\mathcal{O}_{16}+\mathcal{O}_{17}) \right)\\
& +\frac{g'^4}{16\pi^2}\frac{95}{18}\frac{K^{(4)}_{\nicefrac{\mathbf{1}}{\mathbf{2}}}}{\,m_{\nicefrac{\mathbf{1}}{\mathbf{2}}}^4}\left(
-\frac{1}{36}\mathcal{O}_8 +\frac{7}{90}\mathcal{O}_{13} \right)
\,.
\end{split}
\ee
From the KK SM-like vectors without custodial symmetry, we find
\be
\mathcal{L}^{(6)}\supset  -\frac{g^2}{16\pi^2}\frac{1}{48}\frac{K^{(2)}_\mathbf{1}}{m_\mathbf{1}^2}\,\mathcal{O}_{W^3}\,,\ee
\be
\mathcal{L}^{(8)}\supset  \frac{g^4}{16\pi^2}\frac{K^{(4)}_\mathbf{1}}{m^4_\mathbf{1}}\left(
\frac{- 69 \mathcal{O}_{9}-106 \mathcal{O}_{10} + 528 \mathcal{O}_{14} + 228 \mathcal{O}_{15} }{1120}
\right)
\ee
Finally, in presence of custodial symmetry, an additional operator
\be
\mathcal{L}^{(8)}\supset  \frac{g'^4}{16\pi^2}\frac{K^{(4)}_\mathbf{1}}{m^4_\mathbf{1}}\left(
-\frac{5}{32}\mathcal{O}_{8}+\frac{27}{40}\mathcal{O}_{13}
\right)
\ee
is generated. Here, the enhancement factors that take into account the tower of KK resonances are approximately given by $K^{(n)}\approx K^{(n)}_{\rho=0}$ for both fermions and vectors. They are thus approximately given by
$K^{(2)}\approx 1.45$ and $K^{(4)}\approx 1.045$.

\section{Bounds from Higgs and electroweak precision measurements}
\label{constraints}

Let us derive the bounds imposed by electroweak and Higgs precision physics. 
The expected deviations to the $S$ and $T$ parameters ~\cite{Peskin:1990zt,Peskin:1991sw} and to Higgs anomalous couplings in terms of the effective operators of $\mathcal{L}^{(6)}$  are \cite{Dumont:2013wma}
\bea
S&=&\biggl(2s_wc_w\alpha_{WB}+s_w^2 (\alpha_D-2\alpha_{4f})+c_w^2 (\alpha_D'-2\alpha_{4f}')\biggr)\frac{v^2}{\Lambda^2}\,,\nn\\
T&=&\left(-\frac{1}{2}\alpha'_{D^2}+\frac{1}{2}\alpha'_D-\frac{1}{2}\alpha_{4f}'\right)\frac{v^2}{\Lambda^2}\,,
\eea
\bea
a_Z&=&1+\left(\frac{1}{2}\alpha_{D^2}-\frac{1}{4}(\alpha_D-\alpha_{4f})+\frac{1}{4}\alpha'_{D^2}\right)\frac{v^2}{\Lambda^2}\,,\nn\\
a_W&=&1+\left(\frac{1}{2}\alpha_{D^2}-\frac{1}{4}(\alpha_D-\alpha_{4f})-\frac{1}{4}\alpha'_{D^2}\right)\frac{v^2}{\Lambda^2}\,,
\label{basis2aV}
\eea
with  \be
\mathcal{L}_{hV\!V}= a_Z \frac{m_Z^2}{v} h (Z_\mu)^2+a_W \frac{2m_W^2}{v} h (W_\mu)^2\,.
\ee

The presence of IR BKTs modifies the gauge coupling matching and distorts  the gauge propagators. It turns out that for the bulk gauge scenario, the exact predictions for the $S$, $T$ parameters  are 
\be
 S= 2\pi\,f_2(\nu) \left( 1 +  ( r + r' )\, \frac{2+\nu}{3+\nu} \right)\, \frac{v^2}{\tilde{k}^2}\,,\ee
\be
 T= \frac{\pi\,V}{2\,c_w^2}\, f_1(\nu) \left( 1 + \frac{r'}{V} \right)\, \frac{v^2}{\tilde{k}^2}\,,
\ee
and $T=0$ exactly in the custodial case.\footnote{In presence of light exotic fermions there can be important contributions to $S$ and $T$ from loop effects \cite{Grojean:2013qca}. These can be interpreted as arising from operator mixing when performing the RG evolution between the UV and EW scales.} These surprisingly simple  expressions are the result of exact cancellations among  more complicated contributions from the various operators. In particular no such simple expressions can be obtained for the $a_V$ or the $\kappa_i$. One might also check that for the case of a pNGB Higgs, $\nu=0$, and for $r=r'$ the expression for $S$ coincides precisely with the one obtained in Ref.~\cite{Agashe:2004rs}.
Finally, for brane gauge fields, contributions only come at the loop level and are thus much smaller, so that we do not consider them.

The KK gauge fields also induce Higgs anomalous couplings.  
We do not display the full expressions including BKTs, as they are rather lengthy.
For vanishing $r_I$ one gets in the non-custodial case
\be
a_Z\approx 1-\frac{3g^2+2g'^2}{16} V f_1(\nu) \frac{v^2}{ \tilde k ^2}\,\quad a_W\approx 1-\frac{3g^2}{16} V f_1(\nu) \frac{v^2}{ \tilde k ^2}\,.
\ee
In the custodial case one gets \footnote{An important subtlety is that the equality $a_Z= a_W$ is valid only up to terms subleading in the large $V$ expansion. In fact, custodial symmetry in the strong sector ensures only $T=0$, not $a_Z=a_W$.
However one would obtain $a_Z=a_W$ by setting $g'\rightarrow0$. The ungauged hypercharge  limit for which custodial symmetry becomes exact corresponds  in the 5d picture to just switching the $U(1)_R$  UV boundary  condition from Neumann to Dirichlet, and one obtains indeed $a_W=a_Z$ in this case. However, once hypercharge is gauged, the difference $a_Z\neq a_W$ , originating from KK gauge resonances,  is still present.
 }
\be
a_Z\approx a_W\approx 1-\frac{3g^2+3g'^2}{16} V f_1(\nu) \frac{v^2}{ \tilde k ^2}\,.
\ee
In both cases, the $a_W-a_Z$ discrepancy is proportional to $g'^2$. Once including the BKTs, it turns out that in both cases $a_W-a_Z$ grows large for negative values of $r'$  and goes to zero for positive values of $r'$. 
For brane gauge fields, these contributions are zero. 
Finally the radion also generates the new tensorial Higgs couplings
\be
\mathcal{L}_{h\gamma\gamma}= \zeta_\gamma\, h (F_{\mu\nu})^2\,,
\ee
with 
\be
\zeta_\gamma=\frac{\kappa^2}{12}\left(\frac{s^2_w}{r+V}+\frac{c^2_w}{r'+V}\right)\frac{m_h^2\,v}{m_\phi^2\,\tilde k^2}\,.
\ee

For experimental inputs, we use the $S$, $T$ ellipse of the post-Higgs fit \cite{Baak:2012kk} with $U=0$.  
%
%
\be
S=0.05\pm 0.09\,,\qquad T=0.08\pm 0.07
\ee
with a correlation coefficient of $+0.91$.
Constraints on $a_V$ (for $a_W=a_Z$) and the $\zeta_i$ are taken from the global fit of \cite{Dumont:2013wma}. At $95\%$ CL one has 
\be 0.96<a_V<1.21\,,\quad -6.1<\zeta_\gamma v\times 10^{3}<0.9\,, \label{eq:Higgs_bounds}\ee
 the  constraints on other $\zeta_i$ being less stringent for our purpose.


These constraints translate as the following $95\%$ CL bounds on the KK modes and parameters. Consider first vanishing BKTs. From the $S,T$ parameters in the non-custodial case, we obtain\footnote{Slightly more optimistic numbers have been obtained recently in Ref.~\cite{Agashe:2013kxa} for the case of a brane localized Higgs. Note however that these authors quote $99\%$ CL bounds on $\tilde k$ and used an older (pre Higgs-discovery) fit for the $S,T$ ellipse.}
\be
m_{\mathbf{1}}|_{\nu=\infty}> 14.7~\textrm{TeV}\,,\quad m_{\mathbf{1}}|_{\nu=0}>8.1~\textrm{TeV}\,.
\ee
In the custodial case, 
\be
m_{\mathbf{1}}|_{\nu=\infty}>7.7~\textrm{TeV}\,,\quad m_{\mathbf{1}}|_{\nu=0}>6.6~\textrm{TeV}\,.
\ee
From the $a_V$ bounds we have
\be
m_{\mathbf{1}}|_{\nu=\infty}>5.8~\textrm{TeV} \,\quad m_{\mathbf{1}}|_{\nu=0}>3.4~\textrm{TeV}
\ee
in the custodial case. Although the $a_V$ bound is not available in the non-custodial case,  similar numbers are expected.
In case of a bulk Higgs with $\nu=0$, there is therefore no longer a strong motivation to introduce custodial symmetry, as the improvement in the bounds is only marginal. From $\zeta_\gamma$ we also obtain a relatively weak bound on the radion mass,
\be
\tilde k\,m_{\phi}> \kappa \, (221~\textrm{GeV})^2 \,.
\ee

\begin{figure}[t]
\center
\includegraphics[width=7.5cm]{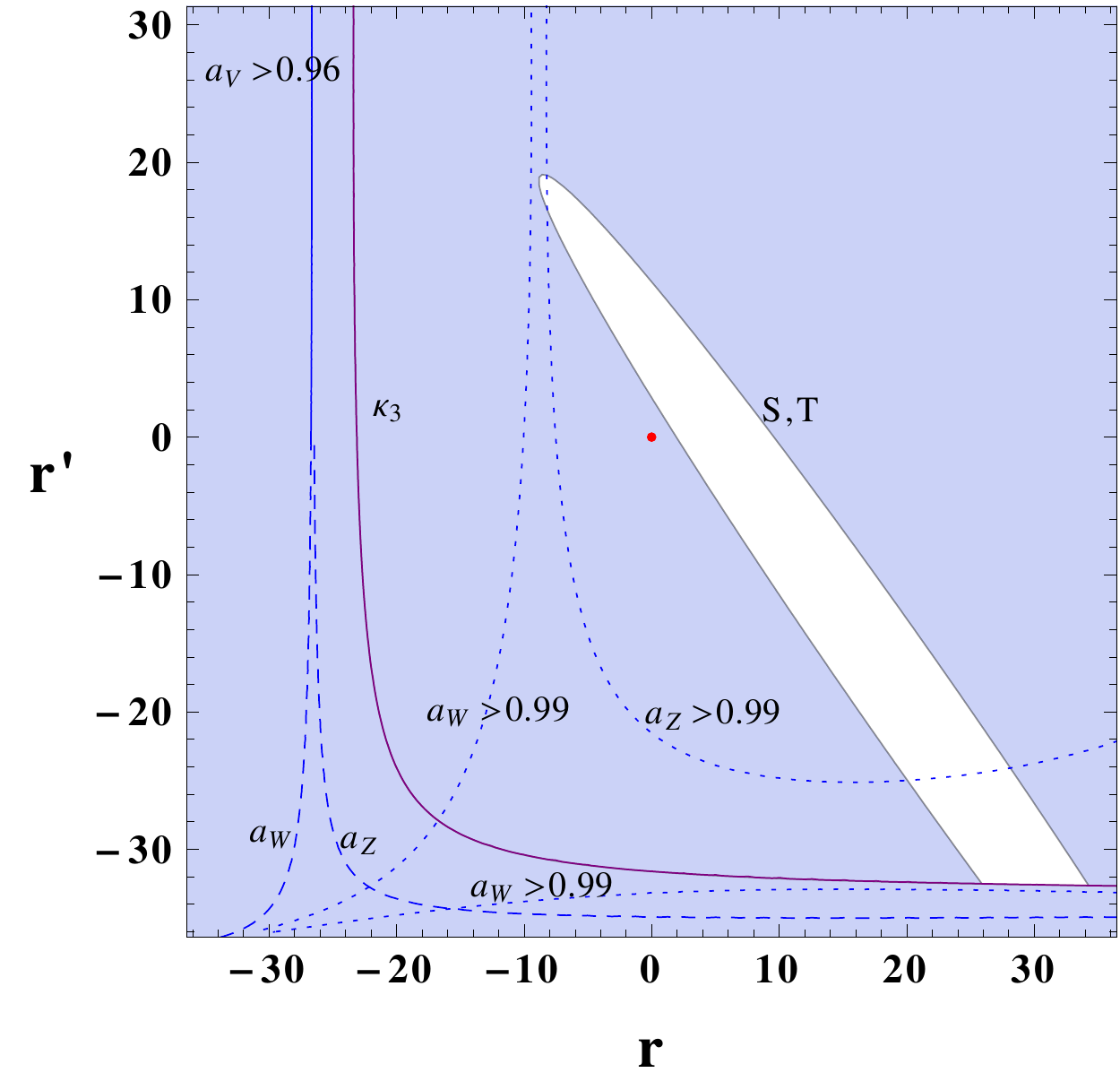}
\includegraphics[width=7.5cm]{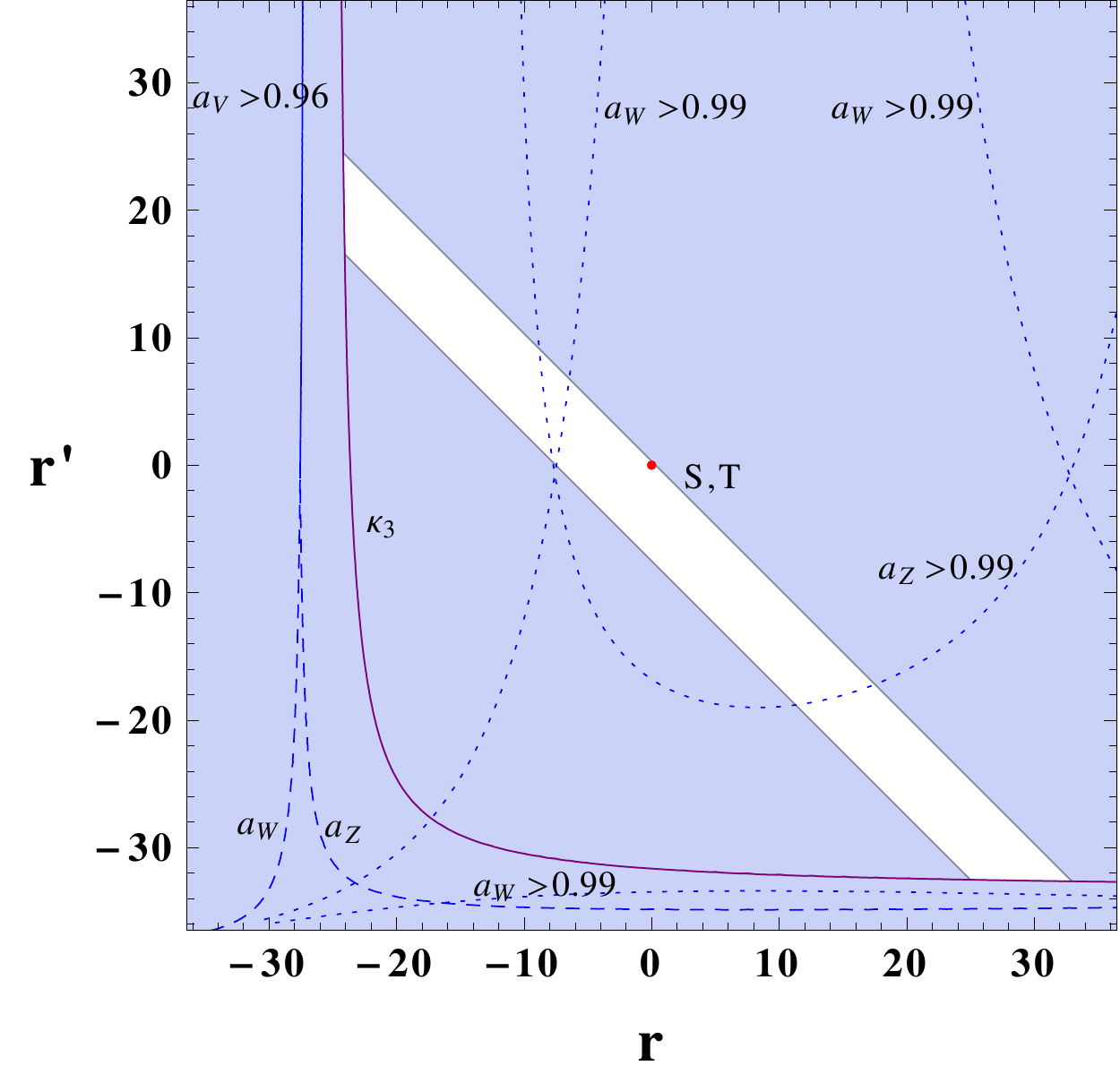}
\caption{
Electroweak, Higgs and gauge precision bounds for $\tilde{k}=3\, \textrm{TeV}$. The regions excluded at $95\%$ CL are shown in blue, the allowed regions are shown in white. The $S,T$ limit is shown in dark, the $a_V$ limit is shown in dark blue and in dashed lines when $a_W\neq a_Z$, and the $\kappa_3$ limit is shown in purple. 
Are also displayed extrapolated limits from more stringent constraints $a_{W,Z}>0.99$, $|\kappa_3|<0.01$.
 The red point denotes the case of vanishing brane kinetic terms.
Left and right panels respectively correspond to non-custodial and custodial cases for a pNGB Higgs ($\nu=0$).
 }\label{fig:STplot}
\end{figure}

Interestingly, the above bounds get relaxed in presence of BKTs. \footnote{For previous attempts of reducing the oblique observables in the presence of BKTs, see Refs.~\cite{Carena:2002dz,Davoudiasl:2002ua}.} The brane and pNGB Higgs cases are similar, we focus on the pNGB Higgs case which is slightly favored.
In both custodial and non-custodial cases a region appears where the contributions to $S,T$ are significantly reduced. The allowed $95\%$ CL regions in the plane $(r,r')$ for  $\tilde{k}=3~\textrm{TeV}$ are shown in Fig. \ref{fig:STplot}. 
We use the $a_V$ constraint when $a_W\approx a_Z$. We also show (dashed lines) how this constraint would apply to $a_W$, $a_Z$ separately. 
It appears that the current constraint on $a_V$, although already stringent for $a_V<1$, is still subleading on most of the parameter space. We also extrapolate how this limit would evolve for $a_{W,Z}>0.99$ instead of $a_V>0.96$. It turns out that such bound would be clearly competitive with respect to the $S,T$ parameters. 
Finally, it is interesting to notice that the favored regions either feature light KK $W_\mu$ or a light KK $B_\mu$, with mass $m_{\mathbf{1}}\approx 0.24\, \tilde{k}$ for respectively $r>0$ and $r'>0$.

%

%
%

%

Here we do not discuss  in detail the limits from direct LHC searches as they are, by the time of this paper,  weaker than the indirect bounds derived above. The most stringent current limits result from KK gluons decaying to top pairs, yielding $m_{\mathbf 1}>2$ TeV \cite{atlasKKgluon}. 
Bounds on KK gravitons are much weaker, with $m_{\mathbf 2}>850$ GeV (at $\kappa=1.0$) for the bulk SM case \cite{atlasbulkKKgrav} 
 and $m_{\mathbf 2}>1$ TeV (at $\kappa=0.1$) for the IR-brane SM case
\cite{Chatrchyan:2012baa}. 
Direct searches are further disfavored at large  $\kappa$ as the KK gravitons are not narrow resonances anymore, while no such problem affects the effective operators. 
A recent study of the discovery potential at the LHC can be found in \cite{Agashe:2013fda}.
We remark that the new indirect bounds for vanishing BKTs derived above most likely push the KK resonances beyond the reach of the LHC, 
 unless some mechanism suppresses the coupling of the Higgs field to electroweak KK modes, for instance  by modifying the geometry in the IR \cite{Fabbrichesi:2008ga,Batell:2008me,Falkowski:2008fz}, reducing in particular the $T$ parameter in non-custodial models \cite{Cabrer:2010si,Cabrer:2011fb,Cabrer:2011vu}.
Another precision observable we do not consider here in detail is the measurement of the $Zb\bar b$ vertex. This quantity is sensitive to the details of the model, including the realization of the custodial sector.\footnote{Bounds comparable to the ones from the oblique parameters (i.e.~$m_{\mathbf 1}\sim4-6$ TeV) can be achieved by localizing the LH bottom quark sufficiently far away from the IR brane \cite{Cabrer:2011qb}.}

\section{Probing a warped extra dimension using anomalous gauge couplings}
\label{results}

\begin{table}
\begin{center}
\begin{tabular}{|c|c|c|c|}
\hline 
Coupling&\multicolumn{2}{c|}{ Bulk gauge fields}&IR  gauge fields\\
 & \hspace{1cm} Non-custodial \hspace{1cm}  & Custodial  &  \\
 \hline
 \hline
$\alpha_D/\Lambda^2$  & $\{0.62,0.46\}/m_{\mathbf{1}}^2$ & $\{0.62,0.46\}/m_{\mathbf{1}}^2$ & $\sim  0$\\
 $\alpha_{D^2}'/\Lambda^2$ & $\{-6.8,-2.3\}/m_{\mathbf{1}}^2$ & $\{0.19,0.14\}/m_{\mathbf{1}}^2$ & $\sim  0$\\
\hline
 \hline
$\kappa_1$ & $\sim 0$ & $\sim 0$ & $\sim 0$\\
$\kappa_{2,3,4}$ & $\{2.88,0.84\}\,v^2/m_{\mathbf{1}}^2$ & $\{-0.38,0.28\}\,v^2/m_{\mathbf{1}}^2$  & $\sim  0$ \\
$\kappa_{5}$ & $\{2.22,0.65\}\,v^2/m_{\mathbf{1}}^2$ & 
$\{-0.29,-0.22\}\,v^2/m_{\mathbf{1}}^2$ 
 & $\sim  0$\\
 \hline
$\lambda^{Z,\gamma}$ & \multicolumn{2}{c|}{$(6.8/m_{\mathbf{\nicefrac{1}{2}}}^2-1.7/m_{\mathbf{1}}^2)\times10^{-4}$ } & $\sim0$\\
\hline
$\eta_1^W$ & \multicolumn{2}{c|}{$\{-0.35,-0.20\}\,\kappa^2\,m_W^2/m_{\mathbf{2}}^4$ } & $-52 \,\kappa^2\,m_W^2/m_{\mathbf{2}}^4$  \\
$\eta_2^W$ & \multicolumn{2}{c|}{$\{0.088,0.049\}\,\kappa^2\,m_W^2/m_{\mathbf{2}}^4$ } & $13\,\kappa^2\,m_W^2/m_{\mathbf{2}}^4$  \\
$\eta_1^Z$ & \multicolumn{2}{c|}{$\{-0.18,-0.098\}\,\kappa^2\,m_Z^2/m_{\mathbf{2}}^4$ } & $-26\,\kappa^2\,m_Z^2/m_{\mathbf{2}}^4$ \\
$\eta_2^Z$ & \multicolumn{2}{c|}{$\{0.044,0.025\}\,\kappa^2\,m_Z^2/m_{\mathbf{2}}^4$ } & $6.5\,\kappa^2\,m_Z^2/m_{\mathbf{2}}^4$ \\
\hline
$\zeta_1^\gamma$  & \multicolumn{2}{c|}{$(0.04\,\kappa^2/(\tilde{k}^2\,m_{\phi}^2)-0.2/m_{\mathbf{\nicefrac{1}{2}}}^4-0.1/m_{\mathbf{1}}^4-3.0\kappa^2/m_{\mathbf{2}}^4)\times10^{-4}$ } & 
$-3.3\,\kappa^2/m_{\mathbf{2}}^4$ \\
$\zeta_1^{\gamma Z}$ &\multicolumn{2}{c|}{  $(0.06/m_{\mathbf{\nicefrac{1}{2}}}^4-0.08/m_{\mathbf{1}}^4) \times 10^{-4}$   } & $\sim0$ \\
$\zeta_1^Z$  & \multicolumn{2}{c|}{$(0.08\,\kappa^2/(\tilde{k}^2\,m_{\phi}^2)-0.07/m_{\mathbf{\nicefrac{1}{2}}}^4-0.7/m_{\mathbf{1}}^4-6.0\kappa^2/m_{\mathbf{2}}^4)\times10^{-4}$ } &
$-6.5\,\kappa^2/m_{\mathbf{2}}^4$  \\
$\zeta_1^W$  & \multicolumn{2}{c|}{$(0.15\,\kappa^2/(\tilde{k}^2\,m_{\phi}^2)-0.2/m_{\mathbf{\nicefrac{1}{2}}}^4-0.7/m_{\mathbf{1}}^4-1.5\kappa^2/m_{\mathbf{2}}^4)\times10^{-4}$ }  &
$-13\,\kappa^2/m_{\mathbf{2}}^4$  \\
$\zeta_2^\gamma$  & \multicolumn{2}{c|}{
$(0.4/m_{\mathbf{\nicefrac{1}{2}}}^4+0.47/m_{\mathbf{1}}^4+
12\,\kappa^2/m_{\mathbf{2}}^4) \times 10^{-4}$ } & 
$13\,\kappa^2/m_{\mathbf{2}}^4$ \\
$\zeta_2^{\gamma Z}$ &\multicolumn{2}{c|}{ $(0.4/m_{\mathbf{\nicefrac{1}{2}}}^4+2.4/m_{\mathbf{1}}^4) \times 10^{-4}$} & $\sim0$ \\
$\zeta_2^Z$  & \multicolumn{2}{c|}{
$(1.4/m_{\mathbf{\nicefrac{1}{2}}}^4+4.4/m_{\mathbf{1}}^4+
24\,\kappa^2/m_{\mathbf{2}}^4) \times 10^{-4}$ } & 
$26\,\kappa^2/m_{\mathbf{2}}^4$ \\
$\zeta_3^Z$  & \multicolumn{2}{c|}{
$(-0.14/m_{\mathbf{\nicefrac{1}{2}}}^4-1.5/m_{\mathbf{1}}^4) \times 10^{-4}$ } & $\sim0$ \\
$\zeta_4^Z$  & \multicolumn{2}{c|}{
$(0.7/m_{\mathbf{\nicefrac{1}{2}}}^4+2.1/m_{\mathbf{1}}^4) \times 10^{-4}$ } &$\sim0$\\
$\zeta_2^W$  & \multicolumn{2}{c|}{
$(1.5/m_{\mathbf{\nicefrac{1}{2}}}^4+5.7/m_{\mathbf{1}}^4+
48\,\kappa^2/m_{\mathbf{2}}^4) \times 10^{-4}$ } & $52\,\kappa^2/m_{\mathbf{2}}^4$ \\
$\zeta_3^W$  & \multicolumn{2}{c|}{
$(-0.3/m_{\mathbf{\nicefrac{1}{2}}}^4-1.1/m_{\mathbf{1}}^4) \times 10^{-4}$ } & $\sim0$ \\
$\zeta_4^W$  & \multicolumn{2}{c|}{
$1.3/m_{\mathbf{\nicefrac{1}{2}}}^4 \times 10^{-4}$ } & $\sim0$ \\
 \hline 
\end{tabular} 
\end{center}
\caption{\label{table_agcs} 
Pattern of the leading  anomalous gauge couplings induced by an $AdS_5$ background with vanishing brane kinetic terms, depending on  gauge fields location and on the presence of custodial symmetry.   The first KK modes are all related to the KK scale $\tilde k$. The first KK gauge field has  $m_{\mathbf{1}}=2.4\, \tilde k$, the first KK graviton has $m_{\mathbf{2}}=3.8 \,\tilde k$. The first SM-like fermion KK mode has $m_{\mathbf{\nicefrac{1}{2}}}=2.4\, \tilde k$ if flat or is heavier otherwise.
The radion mass $m_\phi \ll \tilde k$ is a free parameter.
The number in  brackets correspond respectively to $\nu=\{\infty,0\}$, \ie~respectively to a brane localized Higgs and a pNGB Higgs. 
 The nature of the contribution in the table can be read directly from the index of the mass. 
 The loop contributions from exotic KK fermions can be directly added from Tab. \ref{table_lighttop}.
}
\end{table}


We now consider the anomalous gauge couplings generated from our effective warped extra dimension framework. 
In a first part we discuss the case of vanishing BKTs. The effect of sizeable BKTs will be discussed afterwards. 
The leading contributions  are  summarized in Tab.~\ref{table_agcs}, we now discuss in details its content.

For vanishing BKTs, the contributions from the KK gravitons depend only on the background (here $AdS$) and on the location of gauge fields. They are enhanced by about two orders of magnitude in the brane gauge scenario.
The  contributions from the radion occur only in the bulk gauge  scenario because brane localized gauge fields do not couple to the radion at tree level.
The contributions from KK gauge fields,  KK Higgs, KK fermions  only occur in the bulk gauge scenario as  these modes are absent in the brane gauge scenario.
The KK gauge  contributions depend on custodial symmetry. 
  Finally, the KK Higgs contributions are always negligible with respect to loops of other spin. 
  We only show SM-like KK fermion contributions in Tab.~\ref{table_agcs}.
  They are slightly smaller than the ones from KK gauge fields. 
   Certain fermion KK modes with Neumann-Dirichlet boundary conditions can have a mass $m_{\mathbf{\nicefrac{1}{2}}}\ll \tilde k$, which would enhance significantly their contribution. 
This is precisely how the light exotic resonances of composite Higgs models of Sec.~\ref{CHmodels} appear in the 5d dual description. Their contributions can be added with the help of Tab.~\ref{table_lighttop}.

Let us discuss the  relative size of the contributions to the HDOs. 
Consider first the $\mathcal{L}^{(6)}$ Lagrangian. For bulk gauge fields, the $\mathcal{O}_{D^2}$ and $\mathcal{O}_{D^2}'$ contributions are the largest ones in the non-custodial scenario as they are enhanced by a volume factor $V$. $\mathcal{O}_{D^2}'$ is not enhanced by $V$ in the custodial case (and will actually cancel with $\mathcal{O}_{D}'$, $\mathcal{O}_{4f}'$ within the $T$ parameter). The $\mathcal{O}_{4f}$ contribution is suppressed by  $V$ and is always subleading. 
The  $\mathcal{O}_{D^2}$ does not feed into anomalous gauge couplings but is relevant for Higgs couplings, as discussed in the previous section.
All  $\mathcal{O}_{FF}$ operators can only be loop-generated from renormalizable couplings. However  $\mathcal{O}_{WW}$ and $\mathcal{O}_{BB}$ do
receive tree-level contributions from gravity. These two operators do not contribute to anomalous gauge couplings, but modify Higgs couplings, as discussed in the above Section. In contrast, the  $\mathcal{O}_{WB}$ operator contributes to anomalous gauge couplings, but does not receive contributions from gravity. As other large  contributions are present, we take the computation of $\mathcal{O}_{WB}$ to be beyond the scope of our study, and we choose to neglect it. The dominant contributions to $\mathcal{L}^v_{\rm CGC}$ Lagrangian are therefore   $\mathcal{O}_{D^2}^{'}$ in the non custodial and $\mathcal{O}_{D}$, $\mathcal{O}_{D^2}^{'}$ in the custodial case. 
For the $\mathcal{L}_{\rm CGC}^\partial$ Lagrangian, contributions arise  only from KK $SU(2)_L$ fermions and KK $W$ loops. 

Let us turn to the $\mathcal{L}^{(8)}$ Lagrangian.
Contributions to $\mathcal{L}_{\rm NGC}^v$  come only from KK gravitons. 
For $\mathcal{L}^\partial$,  both KK gravity and KK matter contribute. The radion contribution is small. In the broken phase, it contributes only to the $\zeta_1$ couplings,  because it couples only to the trace of energy-momentum tensors. 
 It turns out that tree-level TeV gravity and loop-level EW processes can be of the same order of magnitude depending on $\kappa^2$. Moreover fermion contributions can be large if one of the KK fermion gets light.  It is thus necessary to keep all the contributions. It is worth noticing that the contribution from the KK gauge loop to $\mathcal{O}_{14}$ is larger by almost one order of magnitude with respect to the ones of the other HDOs. The contributions discussed above will dominate the anomalous couplings and are thus the ones that can be probed in the first place.

Let us now adopt the point of view of anomalous couplings. We first discuss $\mathcal{L}_{\rm CGC}$.
The charged anomalous gauge couplings $\kappa_{2\ldots 5}$ probe the existence of KK gauge modes. In the non custodial case, they are positive, such that one expects a common enhancement of all the cross sections with respect to the SM. In contrast, in the custodial case, they are $<0$, so that a common reduction of the cross-sections is expected. 
They are constrained at LEP, however the bounds from Higgs and electroweak observables derived in the previous section are far more stringent. Using these bounds,  we find that for the bulk gauge fields case, the sensitivity on the $\kappa_i$ needs to reach few per mil to compete with the existing bounds.

The charged anomalous gauge couplings $\lambda^{Z,\gamma}$ probe the existence of $SU(2)_L$ KK gauge modes and of KK fermions.  Although they are loop-generated and thus smaller than the $\kappa_{2\ldots 5}$, one expects the deviation induced by the $\lambda^{Z,\gamma}$ to compete with the  $\kappa_{2\ldots 5}$ in the $v<E<\Lambda$ regime. 
A naive estimate suggests that the effect of these operators in the total cross-section becomes dominant for $E\approx 4\pi v\approx 3$ TeV. However as the effects of this effective coupling grow with the energy, it may be more appropriate to look for deviations in the high-$p_T$  tails of kinematic distributions.  
 The search for these anomalous couplings is particularly appropriate at LHC, which is typically exploring the regime  $v<E<\Lambda$.
 
The $\eta_{1,2}^W$ coupling are sensitive to KK gravitons.
Measuring these couplings seems challenging in ATLAS and CMS because of the SM background. On the other hand FP detectors  like the  ones foreseen for the LHC upgrade  may  help probe these couplings with a smaller background using proton tagging. The $\sigma(\gamma\gamma\rightarrow WW)$ cross-section is enhanced in presence of these couplings \cite{Kepka:2008yx,Chapon:2009hh}. 

Let us turn to $\mathcal{L}_{\rm NGC}$. 
In this work we focus on two-photon neutral couplings as they are forbidden in the SM at tree-level.  The rare processes induced by the anomalous couplings may be probed with high precision using the AFP detector.
The $\eta_{1,2}^Z$ are probing  KK gravitons. Contrary to $\eta_{1,2}^W$, there is no tree-level SM background. They may have a good potential both at ATLAS/CMS and using FP detectors. 
The $\zeta_{1}$ operators receive various contributions from gravity and matter in case of bulk gauge fields, and are rather small. We limit ourselves to a heavy radion, out of reach from direct detection, in order to keep the EFT valid. Its contribution turns out to be small with respect to gravitons. In case of brane gauge fields, the $\zeta_{1,2}$'s  are larger and sensitive to the KK gravitons, while the  $\zeta_{3,4}$'s are vanishing.

Overall,  the states likely to be discovered  by the measurement of anomalous gauge couplings are the KK gauge fields if gauge fields are in the bulk, and the KK gravitons if gauge field are  brane-localized. Fermion contributions are somewhat smaller, although they could be enhanced in presence of  light KK fermions such as those present in the pNGB Higgs scenarios.
 Assuming similar sensitivity for $\eta_{1,2}^{Z,W}$ and the $\zeta_i$'s, the latter might be favored as the LHC typically explores the $v<E<\Lambda$ regime.

\begin{figure}[t]
\center
\includegraphics[width=7cm]{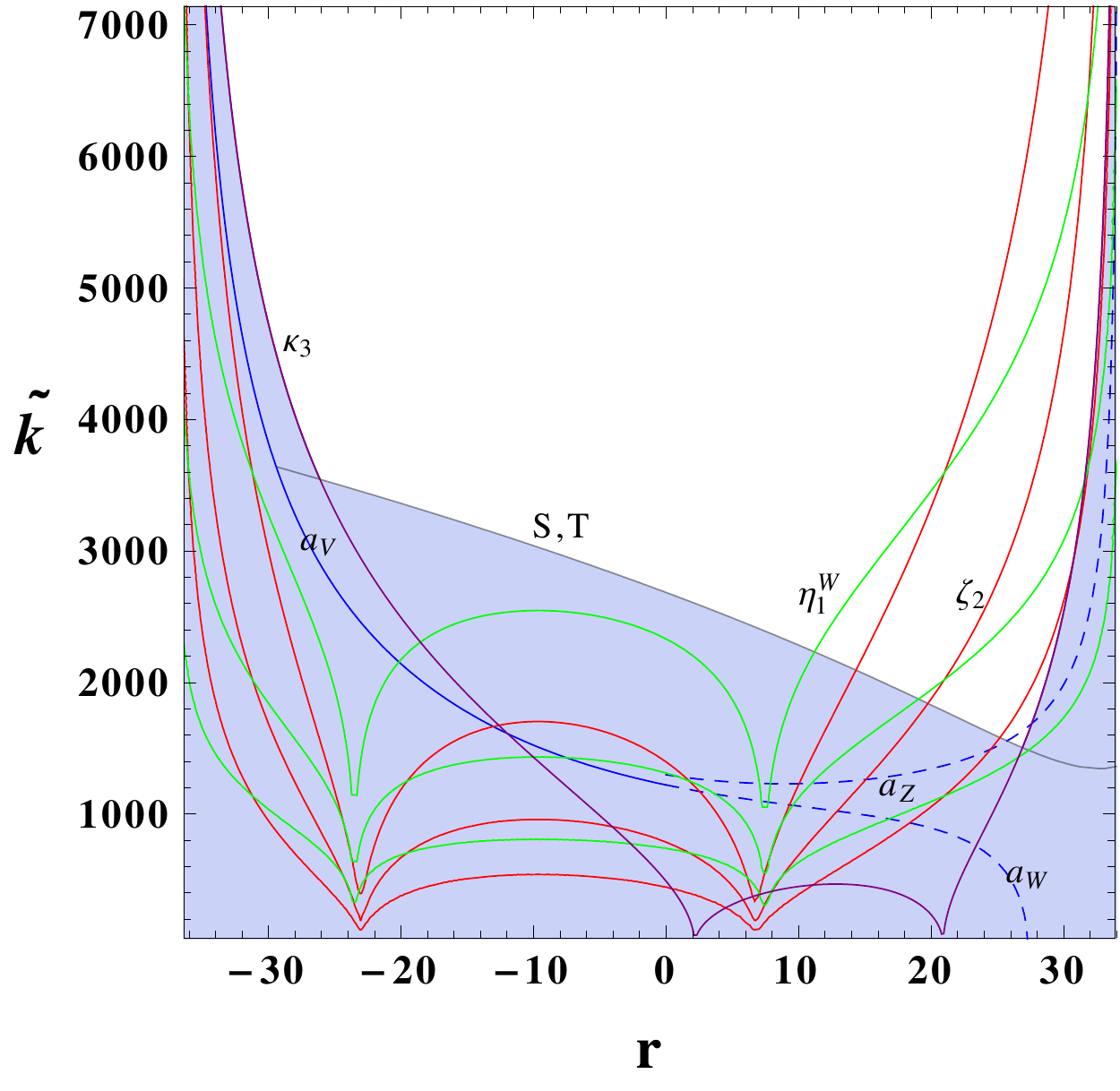}
\includegraphics[width=7cm]{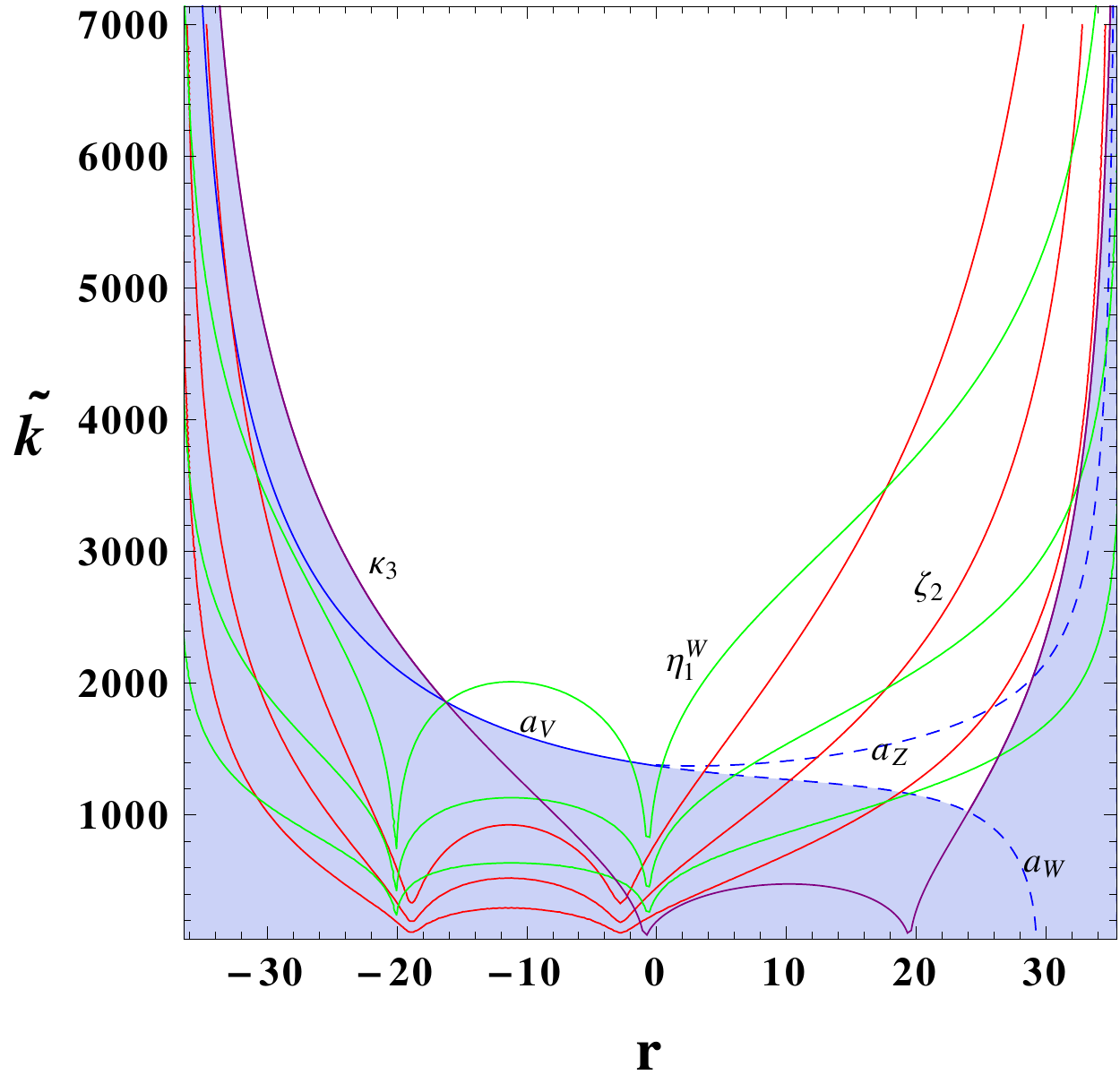}
\caption{
Limits and expected reach in the non-custodial (left) and custodial (right) pNGB Higgs ($\nu=0$) scenarios. 
The reach of $\zeta_2$ (red) and $\eta_1^W$ (green)  from KK gravitons with $\kappa=2$ are shown for sensitivities $\zeta^\gamma_2=(10^{-13},\,10^{-14},\,10^{-15}~\textrm{GeV}^{-4})$ and $\eta^W_1=(10^{-9},\,10^{-10},\,10^{-11}~\textrm{GeV}^{-2})$, from bottom to top.
Color code for other limits is the same as in Fig. \ref{fig:STplot}. 
}\label{fig:AGCplot}
\end{figure}

Let us now consider the case of sizeable BKTs. These BKTs modify the contribution to the $\kappa_{2\ldots5}$ couplings because of the distortion of the gauge propagator. They also modify the gravity contributions to $\eta_i$ and $\zeta_i$ as gravitons
can couple to both bulk and brane components. We focus on the pNGB Higgs case. 
The LEP constraint on $\kappa_{3}(\approx \kappa_2)$ becomes relevant,  the $95\%$ CL interval   translates  as \be0.054<\kappa_{3}<0.021\,,\ee and the limits are displayed on Fig. \ref{fig:STplot}. The $\kappa_3$ bound from LEP, subleading for vanishing BKTs, becomes relevant  for large negative $r,\,r'$.
In both cases, the $a_Z$ grows large for  $r'$ close to $-V$.
It turns out that the most favored regions are the ones with a positive $r$ and a negative $r'$.
The favored slices of parameter space ( corresponding to $r'\approx 7.44 - 1.30\, r$ and $1+\frac{2}{3}( r+ r')=0$ respectively in the non-custodial and custodial cases) are shown  in Fig. \ref{fig:AGCplot}. 
 Interestingly, these regions with relaxed constraints also have enhanced $\eta_i$, $\zeta_i$ anomalous couplings.

The FP detectors provide  a good sensitivity to the $\zeta_{1,2}$, $\eta_1^{W,Z}$ anomalous couplings using proton tagging. 
The latest simulations from the ATLAS Forward Physics collaboration (AFP)
 \cite{royonnewpaper} show an expected sensitivity   \footnote{For this sensitivity a small number of events are obtained. However they have  a large statistical significance as the background is vanishing.  } on $\zeta_{1,2}$ of order $10^{-13}-10^{-14}\,\textrm{GeV}^{-4}$ 
 for $300\,\textrm{fb}^{-1}$. A somewhat older study \cite{Chapon:2009hh} on $\eta_1^{W}$ shows an expected sensitivity of order $10^{-8}\,\textrm{GeV}^{-2}$ for a $5\sigma$ discovery. 
We present the reach from such typical sensitivities in the favored region in Fig. \ref{fig:AGCplot}. 
Are displayed the reaches for $\zeta_2$, $\eta_1^W$, which are slightly better than the other ones.

Finally, regarding the brane gauge scenario,  we find that the expected sensitivities of the FP detectors cited above  lead to observe KK gravitons up to 
\be
m_\mathbf{2}\approx\sqrt{\kappa} \,2.4\, \textrm{TeV}\,
\ee
for the $\eta_1^W$ coupling and 
\be
m_\mathbf{2}\approx \sqrt{\kappa}\,(2.4-4.3)\,\textrm{TeV}\,
\ee
for the $\zeta_1$ coupling. 
 Recall that $\kappa$ can be $O(1)$.  We conclude that the FP detectors can be used to probe KK gravitons in the multi-TeV range in both brane and bulk scenarios.

\section{Conclusion}
\label{conclu}

Anomalous gauge couplings of the Standard Model constitute a rich source of information about the new physics potentially lying above the TeV scale.
These anomalous couplings can be consistently described in an effective Lagrangian approach.
In this paper we aim at going one step further, by mapping actual theories of new physics  onto the effective Lagrangian.
We use an effective parametrization of these theories in order to recast classes of new physics models  in a unified way.


We  first revisit the $SU(2)_L\times U(1)_Y$ effective Lagrangian describing the leading contributions to trilinear and quartic anomalous gauge couplings. We consider all charged gauge couplings, and limit ourselves to neutral couplings involving two photons. We derive the Lagrangian in the $U(1)_{em}$ phase,  which exhibits  the SM anomalous couplings in their form directly usable for phenomenology.
We point out that the whole effective Lagrangian can be split into two pieces $\mathcal{L}=\mathcal{L}^v+\mathcal{L}^\partial$, which are  respectively dominating  in the  regimes of observation  $E<v$ and $v<E<\Lambda$. 
The $\mathcal{L}^\partial$ Lagrangian is particularly relevant from the perspective of TeV  precision physics.

We observe that $\mathcal{L}^\partial$ is  $SU(2)_L\times U(1)_Y$  symmetric and that  all the one-loop perturbative contributions to $\mathcal{L}^\partial$ are only determined by quantum numbers and masses of the spin $0$, $1/2$, $1$ new physics resonances. 
We compute these contributions in a model-independent  form by means of the heat kernel method. The simplified formulas we provide are easily applicable to any model of new physics, as they only require to know how the heavy sector transform under $SU(2)_L\times U(1)_Y$. 

In the framework of  composite Higgs, we derive the pattern of expected deviations for typical $SO(N)$ embeddings of the light top partner which constitute the leading contributions to anomalous couplings. We find contributions to be $O(10^{-4}/m^4_{\Psi})$ for the fundamental $SO(5)$ embedding ($\mathbf{F}_{2/3}$) and $O(10^{-3}/m^4_{\Psi})$ for the symmetric one ($\mathbf{S}_{2/3}$). We discuss how non-pertubative spin-$2$ and dilaton contributions potentially overwhelm a part of the perturbative loop contributions.

In a second part of this work, we study a generic warped extra dimensional framework with $AdS_5$ background. 
By the $AdS$/CFT correspondence,  these theories also provide a 5d weakly coupled description of anomalous gauge couplings induced in the composite Higgs models.
We find  that the anomalous gauge couplings generated in this framework depend mainly on the localization of gauge fields and of brane kinetic terms. The charged anomalous couplings $\kappa_i$ also depend on the assumptions of custodial symmetry. We provide the predictions for all charged and neutral anomalous gauge couplings.

We also study in detail the latest bounds from Higgs precision physics and LEP electroweak precision observables on the first KK gauge mode mass and on the radion mass.  We find the bounds from Higgs couplings to be already  close to compete with those from LEP. For vanishing BKTs, we find KK gauge modes of the RS non custodial scenario to be excluded by $S$, $T$ below $14.7$ and $8.1$ TeV at $95\%$ CL for a brane and   a pNGB Higgs respectively,  and the custodial RS scenario to be excluded below $7.7$ and $6.6$ TeV at $95\%$ CL for a brane and   a pNGB Higgs respectively. All these constraints strongly depend on BKTs, in particular a more favored region appears at negative $r'$. This region features a sizeable discrepancy between $a_W$ and $a_Z$ in both non-custodial and custodial cases, and light KK $W$ are expected.

For the scenario of brane gauge fields, we find that a manifestation of the KK graviton would be the more likely to be first detected. This would most probably happen through the $\eta_{1,2}$ or the $\zeta_{1,2}$ anomalous couplings. As the latter form part of $\mathcal{L}^\partial$, they give enhanced cross sections in the $v<E<\Lambda$ regime and are thus particularly relevant for LHC studies.  
For the scenario of bulk gauge fields, possibly dual to composite Higgs models, we find that the anomalous couplings are most sensitive to the presence of KK gauge modes (\ie~vector resonances) or of KK gravitons (\ie~spin-2 resonances). The KK gauge modes modify the $\kappa$ couplings, while the presence of KK gravitons would be revealed through the $\eta_{1,2}$, $\zeta_{1,2}$ couplings as in the brane scenario. The remaining $\zeta_i$ and $\lambda$ would constitute a probe for KK fermions, if a large enough sensitivity was reached.

%

The future CMS and ATLAS Forward Proton detectors provide ways to probe diphoton anomalous couplings  with high sensitivity, and can be used to test these warped extra-dimensional scenarios.
We find that such precision measurements are complementary of electroweak and Higgs precision physics. 
 Using sensitivity estimations from the AFP collaboration, we find that KK gravitons  can be reached in the multi-TeV mass range in the brane and bulk scenario.

\section*{Acknowledgements}
We would like to thank Eduardo Pont\'on for valuable discussions. We thank O. Kepka, M Saimpert and particularly C. Royon for valuable discussions related to the FP detectors.
GG would like to thank the Funda\c{c}\~ao de
Amparo \`a Pesquisa do Estado de S\~ao Paulo (FAPESP) for financial support, and the International Institute of Physics in Natal for hospitality during part of this project.  SF acknowledges the Brazilian Ministry of Science, Technology and Innovation for financial support, and the Institute of Theoretical Physics of S\~ao Paulo for hospitality.
\\
\\
\\

\newpage

\noindent{\Large\bf Appendix}

\begin{appendix} 

\section{The heat kernel coefficients}

In this appendix we present details of the evaluation of the effective action defined in Eqns.~(\ref{SeffX})--(\ref{Sefff}).
This is conveniently done by rotating to Euclidean space and performing an expansion of the trace of the heat kernel
\be
\tr e^{-t(-D_E^2+X+m^2)}=(4 \pi t)^{-D/2}e^{-t m^2}\sum_{r=0}^\infty \int d^Dx_E\, b^E_{2r}(x)\, t^r
\ee
in powers of $t$ (See Ref.~\cite{Vassilevich:2003xt} for a review). The Gilkey-de Wit coefficients $b^E_n$ depend on the background fields, i.e.~the connection and the field-dependent mass matrix $X$. We are only interested in the UV convergent dimension six and eight operators in a pure Yang-Mills background.
In Minkowsi space, the final result can be given as
\be
\mathcal L_{\rm eff}=(-)^s\frac{1}{32 \pi^2}\left(\frac{1}{m^2}b_6+\frac{1}{m^4}b_8\right)
\ee
The coefficient $b^E_6$ for Yang-Mills background has been computed in Refs.~\cite{Gilkey:1975iq,Fradkin:1983jc}  while $b^E_8$ can be found in Ref.~\cite{Metsaev:1987ju}. 
Converting to Minkowski space,\footnote{Useful relations include $x^0=-ix_E^4$, $x^i=x_E^i$,  $S=iS_E$, $\mathcal L=-\mathcal L_E$ , $V^{ij}= -V_E^{ij}$, $V^{0i}=iV_E^{4i}$.} one finds
\bea
b_6&=&\frac{i}{72}\,\beta\, \tr V_{\mu\nu}V_{\nu\lambda}V_{\lambda\mu}\ \\
b_8&=&\frac{1}{24}\biggl(\gamma_1 \tr V_{\mu\nu}V_{\nu\rho}V_{\lambda_\mu}V_{\rho\lambda}
+\gamma_2 \tr V_{\mu\nu}V_{\nu\rho}V_{\rho\lambda}V_{\lambda_\mu}\biggr.\nn\\
&&\biggl.+\gamma_3 \tr V_{\mu\nu}V_{\mu\nu}V_{\lambda\rho}V_{\lambda\rho}
+\gamma_4 \tr V_{\mu\nu}V_{\lambda\rho}V_{\mu\nu}V_{\lambda\rho}
\biggr)
\eea
up to terms proportional to the equations of motion, $D_\mu V^{\mu\nu}$ that are not needed for our purposes. The $\beta$ and $\gamma_i$ depend on the spin and are given in four dimensions as
\be
\beta^{s=0}=1\,,\qquad \beta^{s=\frac{1}{2}}=4\,,\qquad \beta^{s=1}=4
\ee
\bea
\gamma^{s=0}_i&=&\left(\frac{2}{35},\ \frac{1}{105},\ \frac{17}{210},\ \frac{1}{420}\right)\\
\gamma^{s=\frac{1}{2}}_i&=&\left(-\frac{104}{35},\ -\frac{16}{21},\ \frac{76}{105},\ \frac{64}{105}\right)\\
\gamma^{s=1}_i&=&\left(\frac{344}{35},\ \frac{676}{105},\ -\frac{302}{105},\ -\frac{83}{105}\right)
\eea
For the vector contribution, one has to substract the ghost and add the Goldstone contributions, which amounts to substituting 
$\beta^1\to\beta^1-\beta^0$, and $\gamma^1_i\to\gamma^1_i-\gamma^0_i$. 

We now specialize to the gauge group $G_{SM}=SU(2)\times U(1)$. Labeling the $SU(2)$ representations by their dimension $d$ and the $U(1)$ representations by their hypercharge $Y$, and using the $SU(2)$ identities 
\bea
\tr_d t^{a}t^b &=&C_d\delta^{ab}\nn\\
\tr_d t^at^bt^c&=&\frac{i}{2}C_d\epsilon^{abc}\nn\\
\tr_d t^{a}t^b t^c t^d&=&A_d (\delta^{ab} \delta^{cd}+\delta^{ad}\delta^{bc})+B_d\delta^{ac}\delta^{bd}
\eea
with
\be
A_d=\frac{(d^4-1)d}{240}\,,\qquad 
B_d=\frac{(d^2-9)(d^2-1)d}{240}\,,\qquad
C_d=\frac{(d^2-1)d}{12},,
\ee
we obtain
\bea
b_6&=&-\frac{\beta}{144}\, C_d\, O_{W^3}\nn\\
b_8
&=& \frac{\gamma_3+\gamma_4}{24}\biggl[d\,Y^4\, \mathcal O_8+A_d\, \mathcal O_9+(A_d+B_d)\, \mathcal O_{10}+2Y^2 C_d\, \mathcal O_{11}+4Y^2\, C_d\, \mathcal O_{12}
\biggr]\nn\\
&&+\frac{\gamma_1+\gamma_2}{24} \biggl[d\,Y^4\, \mathcal O_{13}      
+2A_d\, \mathcal O_{14} 
+B_d\, \mathcal O_{15}   
+4Y^2\,C_d\, \mathcal O_{16}
+2Y^2\,C_d\, \mathcal O_{17}
     \biggr]\nn\\
&&+ \frac{\gamma_4}{24}(B_d-A_d)\, (\mathcal O_9-\mathcal O_{10})     
+\frac{\gamma_1}{24}(B_d-A_d)\, (\mathcal O_{14}-\mathcal O_{15})
\eea

\label{app_heatkernel}

 \section{Integrating out a warped extradimension at tree-level}
 \label{app_WED}
 
 \subsection{KK Gravity}
 
 We briefly review the effect of KK resonances of 5d gravity, see Ref.~\cite{Dudas:2012mv} for further details.\footnote{Note that in contrast to Ref.~\cite{Dudas:2012mv} we use here the metric with signature $+----$.}
 We start with the gauge and Higgs Lagrangian
 \be
 \mathcal{L}=-\frac{1}{4}F_{MN}F^{MN}+ D^MH^\dagger D_MH-m^2_H |H|^2\,.
 \ee
 The 5d energy-momentum tensor is defined as 
 \be
 T_{MN}=2\frac{\delta \mathcal{L}}{\delta \gamma^{MN}}-\gamma_{MN}\mathcal{L}\,.
 \ee
 where $\gamma_{MN}=\langle g_{MN}\rangle$ is the 5d background metric given in Eq.~(\ref{metric5d}). 
 It reads 
 \be
 \begin{split}
 T_{MN}^{\rm gauge}&=-F_{MV}F_N^{\,\,V}+\frac{1}{4}\gamma_{MN}F_{PQ}F^{PQ}\\
 T_{MN}^{\rm higgs}&=D_MH^\dagger D_NH +D_NH^\dagger D_MH-\gamma_{MN}(D^PH^\dagger D_PH-m^2_H |H|^2)\,.
 \end{split}
 \ee
 It also contains boundary terms.
 For our purposes we are only interested in gauge and Higgs zero modes.  The zero mode part of the EM tensors is obtained by keeping only
 \be
 A_{M} = \frac{1}{\sqrt{V}}  A_{\mu}^{(0)}+\ldots\,,\quad H = f_H(z)  H^{(0)}+\ldots
 \ee
 where the Higgs profile is 
 \be
 f_H(z)=z^{2+\nu}\,\frac{\sqrt{2+2\nu}}{\sqrt{z_1^{2+2\nu}-z_0^{2+2\nu}}}\,.
 \ee

 The relevant part of the effective action is given as
 \bea
 \mathcal L^{\rm KK - graviton}_{\rm eff}&=&\frac{1}{4\,M^3}\int_{z_0}^{z_1} dz\,dz' \ (kz)^{-3}\,(kz')^{-3}\
  \bar T_{\mu\nu}(x,z)\ G_{\mathbf{2}}(z,z';-\partial_\mu^2)\ \bar T^{\mu\nu}(x,z') \nn\\
   &&  -\frac{1}{2\,M^3}\int_{z_0}^{z_1} dz\,dz' \ (kz)^{-2}\,(kz')^{-2}\
   T_{\mu5}(x,z)\ G_{\mathbf{1}}(z,z';-\partial_\mu^2)\  T^{\mu}_{\ 5}(x,z')
    \nn\\
    &&  +\frac{1}{6\, M^3}\int_{z_0}^{z_1} dz\,dz' \ (kz)^{-1}\,(kz')^{-1}\
  \bar T_{55}(x,z)\ G_{\mathbf 0}(z,z';-\partial_\mu^2)\ \bar T_{55}(x,z') \ ,\nn\\
  \label{Leffgraviton}
 \eea
where we have defined 
 \be
 \bar T_{MN} = 
 T_{MN}-\frac{1}{2}\eta_{MN}T_{\rho}^\rho\,,
 \label{mod}
 \ee
and the propagators $G_{\bf s}$ are defined in App.~A of Ref.~\cite{Dudas:2012mv}.
Eq.~(\ref{Leffgraviton}) is the effective Lagrangian resulting from integrating out the 5d graviton fluctuations in a slice of $AdS_5$.
Due to the presence of zero modes, this Lagrangian is non-local. One zero mode, the 4d graviton, manifests itself as a pole in $G_{\mathbf 2}$ and should be subtracted. The other zero mode, the radion, appears as a pole in $G_{\mathbf 0}$. After stabilization of the extra dimension it acquires a mass and its contribution should be kept. We will take the limit of a light radion, in which there is no effect to leading order besides the non-vanishing mass for the zero mode, which we treat as a free parameter if the theory.

If we are only interested in the zero momentum part of the propagators (no derivatives on the sources) Eq.~(\ref{Leffgraviton}) can be rewritten more explitely in terms of the integrated sources 
 \bea
 \Theta_{\mu\nu}(x,z)&=&\int_{z_0}^z dz'\, (kz')^{-3}\, \bar T_{\mu\nu}(x,z') \ , \nn\\
 \Theta_{\mu }(x,z)&=&\int_{z_0}^z dz' \,(kz')^{-2}\,\bar T_{\mu5}(x,z')\, , \nn \\
 \Theta(x,z)&=&\int_{z_0}^zdz'\, (kz')^{-1}\bar T_{55}(x,z')\, .
 \label{integrated}
 \eea
The effective Lagrangian without the contributinos of the graviton and radion zero modes is then given by
 \bea
 \mathcal L_{\rm eff}&=&\frac{1}{4\,M^3}\int_{z_0}^{z_1} dz\ (kz)^3\,
    \left[\Theta_{\mu\nu}(z)-\Omega_{-2}(z)\,\Theta_{\mu\nu}(z_1)\right]^2\nn\\
 && -\frac{1}{2\,M^3}\left(\int_{z_0}^{z_1} dz\ kz\,\left[\Theta_{\mu }(z)\right]^2
            			-\frac{2k}{\epsilon^{-2}-1}\left[\int_{z_0}^{z_1} dz\ kz\,\Theta_{\mu }(z)
 			\right]^2\right)\nn\\
 &&  +\frac{1}{6\, M^3}\int_{z_0}^{z_1} dz\ (kz)^{-1}\,
         \left[\Theta(z)-\Omega_2(z)\,\Theta(z_1)\right]^2
         \label{final}
 \eea
 with the definition
 \be
 \Omega_p=\frac{z^p-z_0^p}{z_1^p-z_0^p}\,.
 \ee
For the radion, one simply obtains
\be
\mathcal L_{\rm eff}=\frac{\epsilon^2}{3 M_{Pl}^2}\ \frac{1}{m_\phi^2}\ \left[\Theta(x,z_1)\right]^2\,.
\ee

For reference, we explicitely write down the relevant sources as functions of the (canonically normalized) bosonic zero modes
\bea
\Theta_{\mu\nu}(x,z)&=&
\frac{\log(kz)}{V+r_I}\left(-F^I_{\mu\rho} F^{I\,\rho}_{\ \nu}+\frac{1}{4}\eta_{\mu\nu}F^I_{\rho\sigma} F^{I\,\rho\sigma}\right)\nn\\&&
+\Omega_{2(\nu+1)}\left(D_\mu H^\dagger D_\nu H+D_\nu H^\dagger D_\mu H\right)
-2(1+\nu)(2+\nu)\,\tilde k^2\,\frac{z^{2\nu}}{z_1^{2\nu}}\eta_{\mu\nu}|H|^2
\nn\\
\Theta_{\mu\nu}(x,z_1)&=&-F_{\mu\rho}^I F^{I\,\rho}_{\ \nu}+\frac{1}{4}\eta_{\mu\nu}F^I_{\rho\sigma} F^{I\,\rho\sigma}
+D_\mu H^\dagger D_\nu H+D_\nu H^\dagger D_\mu H
-m^2\eta_{\mu\nu}|H|^2\nn\\
\Theta(x,z_1)&=&-\frac{1}{8(V+r_I)}\epsilon^{-2}F^I_{\rho\sigma} F^{I\,\rho\sigma}+2\epsilon^{-2}m^2|H|^2
\label{sources}
\eea
Here, $m^2=2(1+\nu)[(2+\nu)\tilde k+\epsilon\,m^{\rm brane}_H]\tilde k$ is the effective mass term in the Higgs potential, i.e., the physical Higgs mass (assuming a quartic potential) is $m_h^2=-2m^2$\,.

  \subsection{KK gauge}
  
The general matter-gauge interactions have the form 
\be
S\supset -g_5 \int d^5x\, \sqrt{g} A_M^I(x,z)\sum_X \mathcal{J}^{M,I}_{X}(x,z) \,.
\label{gauge1}
\ee
We need to keep only the zero mode part of the 5d currents,
 \be
\mathcal{J}^{\mu}_{X}(x,z)= (kz)^2 \omega_X(z)  J^{\mu}_{X}(x)+\ldots
 \ee
with \be
\omega_H=(kz)^{-3}  f_H(z)^2\,,\quad \omega_f=(kz)^{-4} f_f(z)^2\,.
\ee 
Moreover, as we are only interested in the zero-momentum propagators, it is convenient to define 
\be
\Omega_X=\int_{z_0}^z dz\,\omega_X\,.
\ee
i.e.~$\Omega_H=\Omega_{2(\nu+1)}$ and $\Omega_f=\Omega_{1-2c_f}$, where $c_f$ is the fermion localization parameter. For the purpose of this paper we can take the oblique approximation $\Omega_f\approx \Omega_{-\infty}\equiv 1$.
  The 4d effective Lagrangian is then
 \be
 \begin{split}
\mathcal{L}_{\rm eff} =&-\frac{g_5^2}{2}\left( a_{HH} J_H^{W}\cdot J_H^{W}+2a_{Hf} J_H^{W}\cdot J_f^{W}+a_{ff} J_f^{W}\cdot J_f^{W}   \right)\\
&-\frac{g_5'^2}{2}\left( a_{HH} J_H^{Y}\cdot J_H^{Y}+2a_{Hf} J_H^{Y}\cdot J_f^{Y}+a_{ff} J_f^{Y}\cdot J_f^{Y}   \right)\,.
\label{gauge2}
\end{split}
 \ee
Notice that we work here with mostly minus signature, while the signature is mostly plus in \cite{Cabrer:2011fb}, hence the overall minus sign in Eqns.~(\ref{gauge1}) and (\ref{gauge2}). The SM gauge bosons have  Neumann-Neumann ($++$) boundary conditions and the corresponding coefficients read:
 \be
a_{XY} =\int dz \ kz\ \left(\Omega_X- \Omega_0  \right)\left(\Omega_Y- \Omega_0  \right)\,,\qquad (+\!+{\rm \ gauge\ field)}
\label{aXY}
 \ee
where $\Omega_0(z)=\log(kz)/V$. 
Finally, the $\mathcal{O}_{D^2}$, $\mathcal{O}_{D^2}'$  are related to $(J_H^{Y})^2$,  $(J_H^{W})^2$ as
\be
(J_H^{Y})^2= \frac{1}{2} \mathcal{O}_{D^2}+\mathcal{O}_{D^2}'+\frac{1}{4}( |H|^2 H^\dagger D^2 H+h.c.)  \,,
\ee
 \be
(J_H^{W})^2= \frac{3}{2} \mathcal{O}_{D^2}+ \frac{1}{4}( |H|^2 H^\dagger D^2 H+h.c.) \,.
\ee
In case there are additional gauge bosons from extended bulk gauge sectors one has to apply the expressions for mixed boundary conditions. In this case, the $\alpha_{XY}$ have to be replaced by
\begin{align}
a_{XY}&=\int dz\ kz\ \Omega_X\,\Omega_Y\,,&(+\!-{\rm \ gauge\ field)}\\
a_{XY}&=\int dz\ kz\,\left(\Omega_X-1\right)\left(\Omega_Y-1\right)\,,& (-\!+{\rm \ gauge\ field)}
\end{align}
for the indicated (UV,IR) boundary conditions respectively. 
We will not need the $--$ BC case for this paper, but expressions can be found in Ref.~\cite{Cabrer:2011fb}.
Notice that for the custodial case, there is the relation
\be
\left(J_H^{W^1_R}\right)^2+\left(J_H^{W_R^2}\right)^2=\mathcal O_{D^2}-O'_{D^2}
\ee
Adding brane kinetic terms (BKT) only modifies the $++$ contribution $a_{XY}$ ($+-$ ($-+$) fields do not couple to the IR (UV) brane, while the $-+$ and $+-$ contributions to $a_{XY}$ are unchanged because the respective propagators are only modified at higher order in the momentum expansion). The necessary modification of $a_{XY}$ turns out to be remarkably simple: One just needs to generalize $\Omega_0$ to 
\be
\Omega_I=\frac{\log(kz)+r^I_0}{V+r^I_0+r^I_1}\,,
\ee
in the expression Eq.~(\ref{aXY}). This can be shown by carefully including BKTs in the calculation of the momentum-less propagator of Ref.~\cite{Cabrer:2011fb}. In particular, one can see the limit of infinite UV BKT ($r_0\to\infty$) reproduces 
 the $-+$ result ($\Omega_I=1$), while for infinte IR BKT ($r_1\to\infty$) one finds the $+-$ result ($\Omega_I=0$).
 
 \end{appendix}

\end{document}